\documentclass[aps,prx,twocolumn,footinbib,showpacs,showkeys,nobalancelastPage]{revtex4-1}
\usepackage{graphicx}
\usepackage{indentfirst}
\usepackage{physics}
\usepackage{braket}
\usepackage{float}
\usepackage{amsmath,amssymb}
\usepackage{autobreak}
\usepackage{upgreek}
\usepackage{verbatim}
\usepackage{epstopdf}
\usepackage{CJK}
\usepackage{esint}
\usepackage{color}
\usepackage[T1]{fontenc}
\usepackage{subfigure}
\usepackage{amsfonts}
\usepackage{footmisc}
\usepackage{scrextend}
\usepackage{multirow}
\usepackage[hyperfootnotes=false]{hyperref}

\usepackage[english]{babel}
\usepackage{url}
\usepackage{bm}
\usepackage{hyperref}
\definecolor{darkblue}{rgb}{0,0,0.5}
\hypersetup{
colorlinks=true,
linkcolor=black,
filecolor=blue,
citecolor=darkblue,  
urlcolor=black,
}

\usepackage{enumitem,kantlipsum}
\usepackage{refcount}

\newcommand{\calE}{{\cal E}}

\newcommand{\calL}{{\cal L}}
\newcommand{\calN}{{\cal N}} 

\newcommand{\calG}{{\cal G}}

\newcommand{\calR}{{\cal R}}

\newcommand{\1}{^{(1)}}

\def\be{\begin{equation}}
\def\ee{\end{equation}}
\def\bl{\begin{align}}
\def\el{\end{align}}
\def\ba{\begin{eqnarray}}
\def\ea{\end{eqnarray}}
\usepackage{bm}

\begin{document}

\title{Entanglement formation in continuous-variable random quantum networks}
\author{Bingzhi Zhang$^{1,2}$}
\author{Quntao Zhuang$^{2,3}$}
\email{zhuangquntao@email.arizona.edu}
\affiliation{
$^1$Department of Physics, University of Arizona, Tucson, AZ 85721, USA
\\
$^2$Department of Electrical and Computer Engineering, University of Arizona, Tucson, AZ 85721, USA
\\
$^3$James C. Wyant College of Optical Sciences, University of Arizona, Tucson, AZ 85721, USA
}
\date{\today}

\begin{abstract}
Entanglement is not only important for understanding the fundamental properties of many-body systems, but also the crucial resource enabling quantum advantages in practical information processing tasks.
While previous works on entanglement formation and networking focus on discrete-variable systems, light---as the only travelling carrier of quantum information in a network---is bosonic and thus requires a continuous-variable description in general. In this work, we extend the study to continuous-variable quantum networks. By mapping the ensemble-averaged entanglement dynamics on an arbitrary network to a random-walk process on a graph, we are able to exactly solve the entanglement dynamics and reveal unique phenomena. 
We identify squeezing as the source of entanglement generation, which triggers a diffusive spread of entanglement with a parabolic light cone. 
The entanglement distribution is directly connected to the probability distribution of the random walk, while the scrambling time is determined by the mixing time of the random walk.
The dynamics of bipartite entanglement is determined by the boundary of the bipartition; An operational witness of multipartite entanglement, based on advantages in sensing tasks, is introduced to characterize the multipartite entanglement growth.
A surprising linear superposition law in the entanglement growth is predicted by the theory and numerically verified, when the squeezers are sparse in space-time, despite the nonlinear nature of the entanglement dynamics. 
We also give exact solution to the equilibrium entanglement distribution (Page curves), including its fluctuations, and found various shapes dependent on the average squeezing density and strength. 
\end{abstract}

\maketitle


\section{Introduction}

Quantum information science has brought to us capabilities to enhance the performance of computing~\cite{Shor_1997}, sensing~\cite{Giovannetti2004} and communication~\cite{Bennett2002,gisin2007quantum}, through entangling local or distant processing nodes. Therefore, a quantum network~\cite{kimble2008quantum} that enables entanglement establishment is important for achieving the promised quantum advantages.
The study of entanglement formation and quantum information scrambling has been fruitful in complex systems such as random quantum networks~\cite{biamonte2019complex,brito2019statistical,acin2007entanglement}, and circuits~\cite{nahum2017quantum,nahum2018operator,Keyserlingk:2018aa,Khemani:2018aa,Rakovszky:2018aa}, many-body systems~\cite{kim2013ballistic,luitz2017information,huang2017out,chen2017out,fan2017out,gopalakrishnan2018hydrodynamics,you2018entanglement,banerjee_2017,patel2017quantum,patel2017quantum_prx}, models of holography~\cite{sachdev1993,kitaev2015simple,gu2017local,kitaev2018soft} and quantum gravity~\cite{hayden2007black, hosur2016chaos, Yoshida:2017aa, gao2017traversable, maldacena2017diving,sekino2008fast,maldacena2016bound,shenker2014black,Lashkari13, Roberts:2014isa,piroli2020random,agarwal2020toy,liu2020dynamical}. Universal scaling laws and dynamical models of entanglement formation has been established, based on nonlinear surface growth models~\cite{nahum2017quantum,nahum2018operator}. Recently, experimental probing~\cite{landsman2019verified,fan2017out,garttner2017measuring} of scrambling is also made possible; from the quantum network perspective, protocol designs~\cite{chakraborty2019distributed,vardoyan2019performance,pant2019routing} for entanglement establishment has also been a recent focus.

The above works, whether on the basic understanding of scrambling or practical design of networking, mainly focus on entanglement in discrete-variable (DV) systems, which is natural for computing. However, as quantum networks inevitably utilize light as the carrier of quantum information in transmission, the bosonic nature of light makes it necessary to consider entanglement in a continuous-variable (CV) description. Moreover, various applications in the photonic or microwave domain, including universal quantum computing based on cluster states~\cite{menicucci2006universal}, quantum illumination~\cite{tan2008quantum,zhuang2017optimum,Zheshen_15}, quantum reading~\cite{pirandola2011quantum}, distributed sensing~\cite{zhuang2018distributed,zhuang2019physical,guo2020distributed,xia2019entangled} and entanglement-assisted communication~\cite{shi2019practical,guha2020infinite}, require CV entanglement in the form of Gaussian states~\cite{Weedbrook_2012}. 
In this regard, noiseless linear amplifiers~\cite{seshadreesan2018continuous} and novel error correction codes~\cite{noh2019encoding,zhuang2019distributed} provide initial tools for CV networking, and an out-of-time-order correlator (OTOC) has revealed a unique squeezing-dependent butterfly-velocity of operator spreading~\cite{zhuang2019scrambling}.

\begin{figure}
    \centering
    \subfigure{
    \includegraphics[width = 0.45\textwidth]{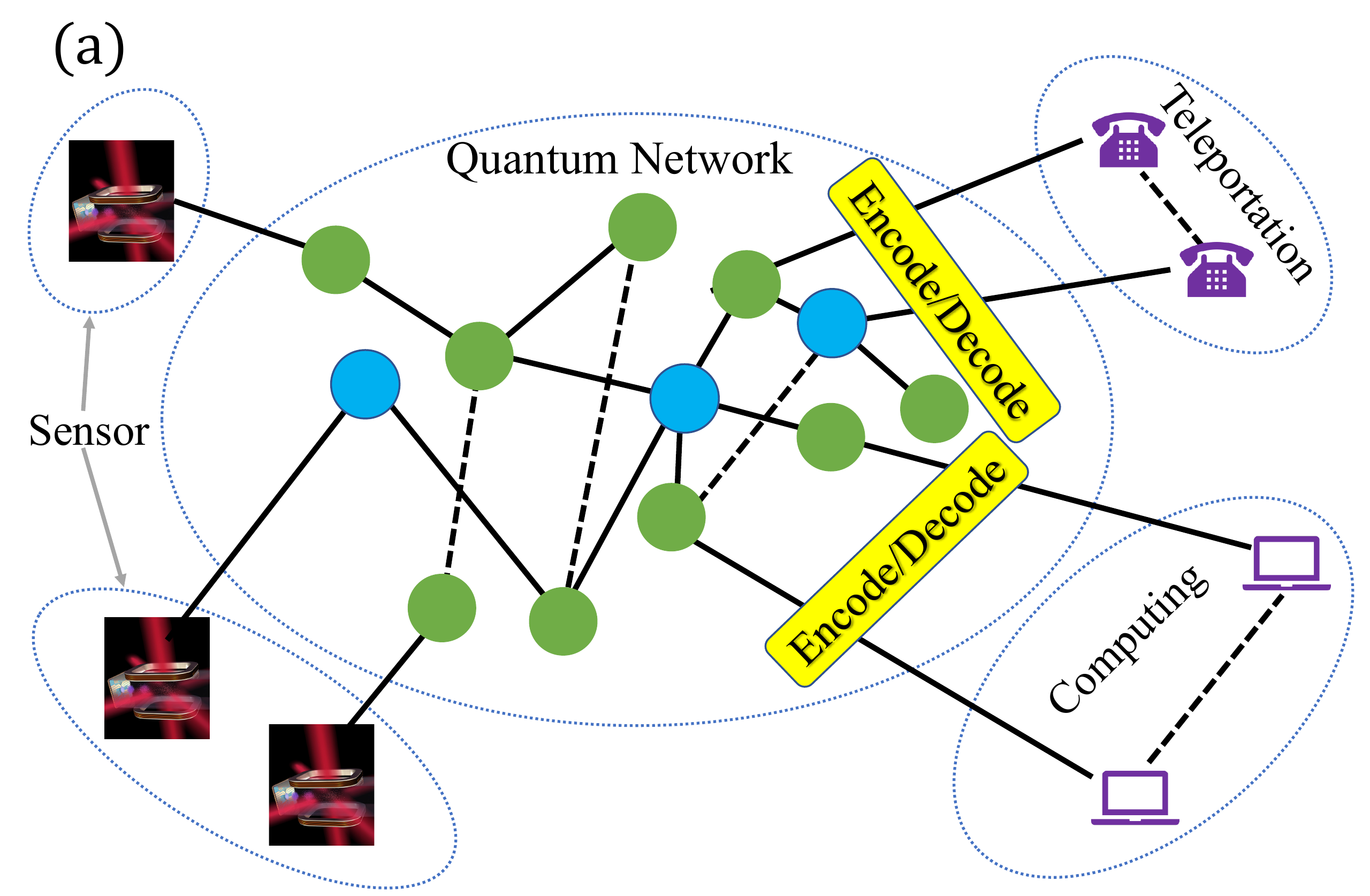}
    }
    \subfigure{\includegraphics[width = 0.45\textwidth]{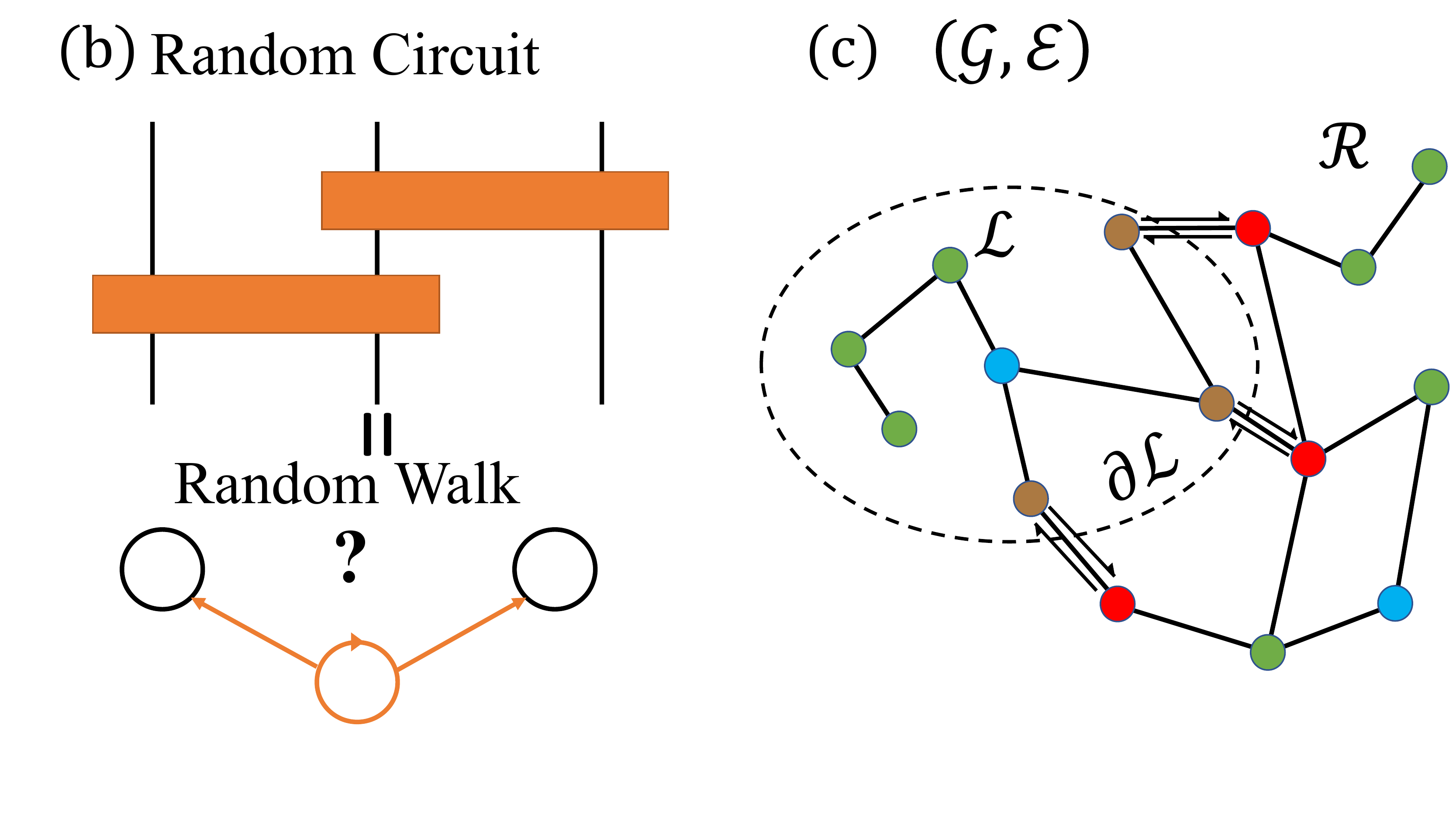}}
    \caption{(a) Schematic of a quantum network. Through entanglement distribution, the nodes get entangled to enhance communication, sensing and computation tasks. Cyan and green nodes represent nodes with or without squeezers. Solid and dashed lines represent quantum and classical links. (b) Random unitary gates correspond to random walk on a graph. (c) A general graph $(\calG,\calE)$, where $\calG$ denotes all the vertices and $\calE$ denotes all edges. Time evolution of the entanglement entropy between the subsystem $\calL$ and the rest $\calR$ depends on the random walk on the boundaries $\partial\calL$. Brown vertices are the inner boundaries $\partial \calL^-$ (vertices inside $\calL$ but connected to $\calR$) and red vertices are the outer boundaries $\partial \calL^+$ (vertices outside $\calL$ but connected to $\calL$).
    \label{fig:general_graph}
    }
\end{figure}

In this paper, we study quantum information scrambling in CV quantum networks (see Fig.~\ref{fig:general_graph}) focusing on the entanglement formation dynamics. 
Inspired by the classical statistical theory of complex networks~\cite{barabasi1999emergence,watts1998collective,barabasi1999mean}, we consider random quantum networking protocols to enable analytical solutions, through a mapping to a random-walk process on graph; at the same time, random protocols are expected to reveal typical and universal characteristics. Our results apply to quantum networks on general graphs, therefore provide a foundation for the statistical theory of complex quantum networks.

We provide an analytical formula connecting the entanglement entropy to weights in the passive linear optical transforms, therefore establishing a mapping between ensemble-averaged entanglement dynamics to the probability evolution of a random-walk process on a general graph (see Fig.~\ref{fig:general_graph}). 
The change of the entanglement entropy $S(\calL,t)$ of a subsystem $\calL$, similar to the random walker's probability in subsystem $\calL$, is determined by the boundary $\partial\calL$. Moreover, we also solve the fluctuations of the entanglement entropy.

In discrete space-time, the ensemble-averaged weights dynamics can be described by a Markov chain, with the transition matrix determined by the graph connectivity. Going into continuous space-time, we can derive simple diffusive partial differential equations (PDEs) to describe the entanglement evolution. In both cases, the model is completed with the analytical formula connecting weights to entanglement entropy. Alternatively, a phenomenological model coupling an epidemiology equation with a nonlinear diffusion can directly capture the entanglement dynamics.

To go beyond the bipartite characterization of entanglement through entanglement entropy, we also give an operational witness of multipartite entanglement. This witness is directly connected to entanglement's advantage over classical correlations in distributed sensing protocols~\cite{zhuang2019distributed}. Maximum values of the entanglement witness are achieved towards the late time, therefore verifying the full scrambling of the entire network.

Through the mapping between quantum dynamics and random walk, we also connect the scrambling time---the time it takes for the entire system to be maximally entangled---directly to the mixing time of the random walk. 
Moreover, at infinite time, the equilibrium entanglement distribution---analog to the Page curve in DV systems~\cite{Page93,nakagawa2018universality,fujita2018Page}---can be solved analytically from the stationary state of the random walk. Surprisingly, the Page curve is independent of the topology of the network, as long as the graph is connected. And in general it depends on two statistical properties of the quantum network---the squeezer's density and the average squeezing strength. Interestingly, a small subsystem can almost get close to the maximum entanglement entropy, while in DV systems, half of the system size is necessary.

While our theory works for general graphs, we also apply to networks respecting `locality' of interactions---$D$-dimensional Cartesian graphs where links only exist between nearest neighbors (see Figs.~\ref{fig:schematic_1D} and~\ref{fig:schematic_2D}). In this regard, we identify a diffusive entanglement light cone at the early time, which divides the regions with almost no entanglement and regions with substantial entanglement. After the entanglement light cone reaches each node, there is a period of entanglement sudden growth, where the entanglement entropy quickly gets close to its equilibrium value. In the end, there is a long period of saturation, determined by the mixing time $\sim M^2$ quadratic in the length $M$ on each dimension.

Our theory framework provides a unique complement to the DV counterparts~\cite{nahum2017quantum,nahum2018operator,Keyserlingk:2018aa,Khemani:2018aa,Rakovszky:2018aa,you2018entanglement}.
And our results provide insights into not only CV quantum networks being engineered, but also quantum information scrambling in various physical systems, as any form of bosonic radiation is intrinsically CV. 

The paper is organized as the following. In Sec.~\ref{sec:intro_Gaussian}, we specify the model in details and give a more specific overview of results; in Sec.~\ref{sec:statistical_theory}, we present the statistical theory for the single-squeezer case based on a mapping to random-walk dynamics; in Sec.~\ref{sec:multiple_squeezers}, we generalize the single-squeezer results to the general case through providing a linear superposition law.

\section{Quantum networks: modelling and main results}
\label{sec:intro_Gaussian}

Our overall goal is to characterize generic entanglement formation dynamics towards equilibrium in a CV quantum network (see Fig.~\ref{fig:general_graph} for a schematic). In general, a quantum network can have complicated topology, which makes the problem difficult. Moreover, each node can possess multiple optical modes, and perform local operations coordinated by classical communication to entangle them. 
\begin{figure}
\centering
\includegraphics[width = 0.475\textwidth]{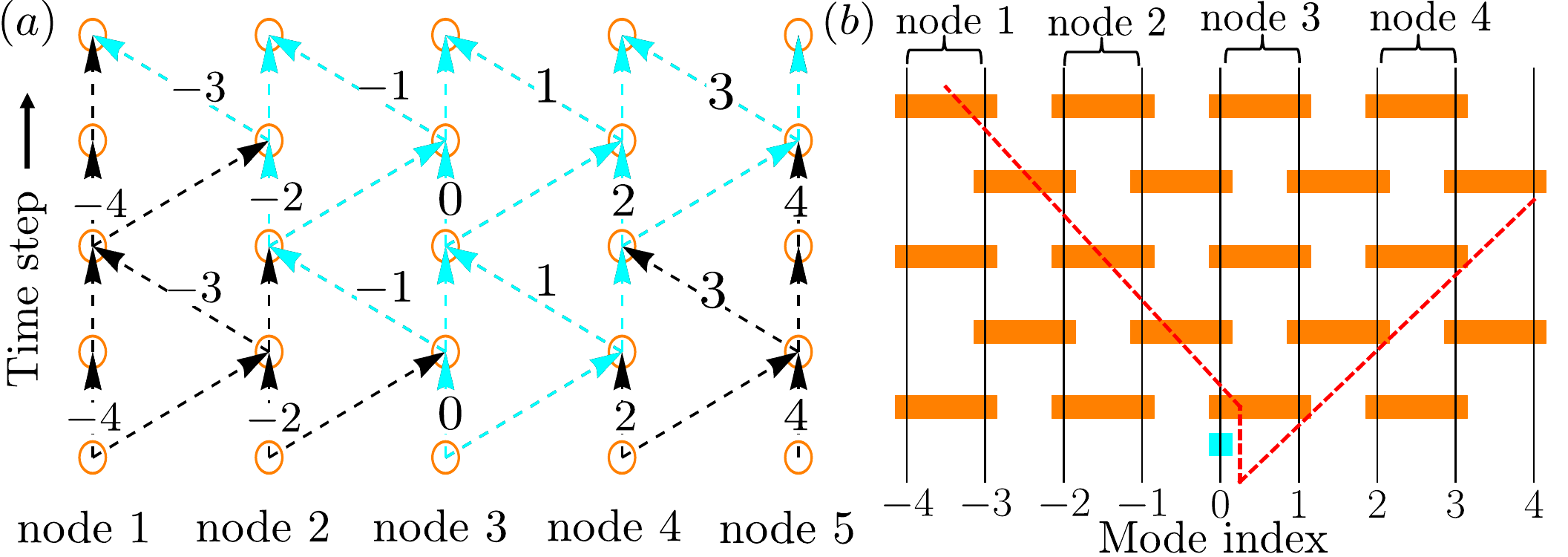}
\caption{Schematic of the (a) 1-D hopping quantum network and (b) its corresponding 1-D local circuit. In total, $M=2N+1$ modes are utilized in $N+1$ network nodes ($N=4$ case is plotted). We index the $M=2N+1$ bosonic modes by integers $x\in [-N,N]$. For convenience, we also introduce a scaled coordinate $\tilde{x} = x/2N\in[-0.5, 0.5]$. Here we exemplify the notation: the connection matrix $E_{x,x^\prime}=\delta_{|x-x^\prime|-1}$; neighbors $\calN(x)=\{x-1,x,x+1\}$; it is convenient to consider the left part of the system $\calL=[-N,x]$, and the right part of the system $\calR=[x+1,N]$.
In (b) the cyan square is a single-mode squeezer, while the orange rectangles are the 2-mode beamsplitters (combing phase shifters). The red dashed line represents the light cone from the center. In (a) each empty circle denotes a local mixing operation by beamsplitters and the dashed arrows indicate the transmission of optical modes in the network. The cyan lines indicate the light cone starting from the vertex with the single-mode squeezer.
\label{fig:schematic_1D}
}
\end{figure}
Considering the optical modes, we can reduce a general entanglement generation protocol to a quantum circuit on a graph, as we illustrate in an one-dimensional (1-D) hopping quantum network in Fig.~\ref{fig:schematic_1D} (a). To establish entanglement, each node performs the following protocol repetitively: it receives a light mode from a neighbor, which gets entangled with a stored mode through a local unitary; then, it sends out a mode to another neighbor and stores one mode locally. For simplicity, the nodes send light to the left and right neighbors alternatively in even and odd steps. If we focus on the dynamics of the optical modes, the above protocol reduces to a 1-D local circuit in Fig.~\ref{fig:schematic_1D} (b), where local gates apply alternatively on the light modes~\cite{zhuang2019scrambling}.

The transmission links in a quantum network are in general lossy. To cope of loss, error correction~\cite{noh2019encoding} can be applied in each link transmission. On the physical layer, this means including additional components that seemingly complicate the analyses. However, on the logical layer, up to some small residual errors from imperfect error correction, the state being protected is identical to the state being generated in a lossless quantum network, as demonstrated in Ref.~\cite{zhuang2019distributed} for sensing purposes. Therefore, we start with the lossless case.


With the mapping between quantum networks and quantum circuits in mind, we specify the set-up of the circuits on an arbitrary (undirected) graph (see Fig.~\ref{fig:general_graph}(c)). In general, the topology can be described by an un-directed graph $(\calG, \calE)$, where $\calG$ denotes the set of all vertices, each described by a coordinate system $\bm x$~\cite{Note3}. The set of edges $\calE$ can be described by a generalized connection matrix $E_{\bm x, \bm x^\prime}$. When $E_{\bm x, \bm x^\prime}=1$, the vertices $\bm x, \bm x^\prime$ are connected by an edge $\overline{\bm x \bm x^\prime}$, zero when not connected. For simplicity, we write the set of vertices that are directly connected to $\bm x$ (neighbors) as $\calN(\bm x)$. We are interested in the entanglement between a set of vertices $\calL$ and the rest $\calR=\calG \backslash \calL$. In Fig.~\ref{fig:schematic_1D}(b), we give a $1$-D example of the notations.

\begin{figure}
\centering
    \includegraphics[width = 0.45\textwidth]{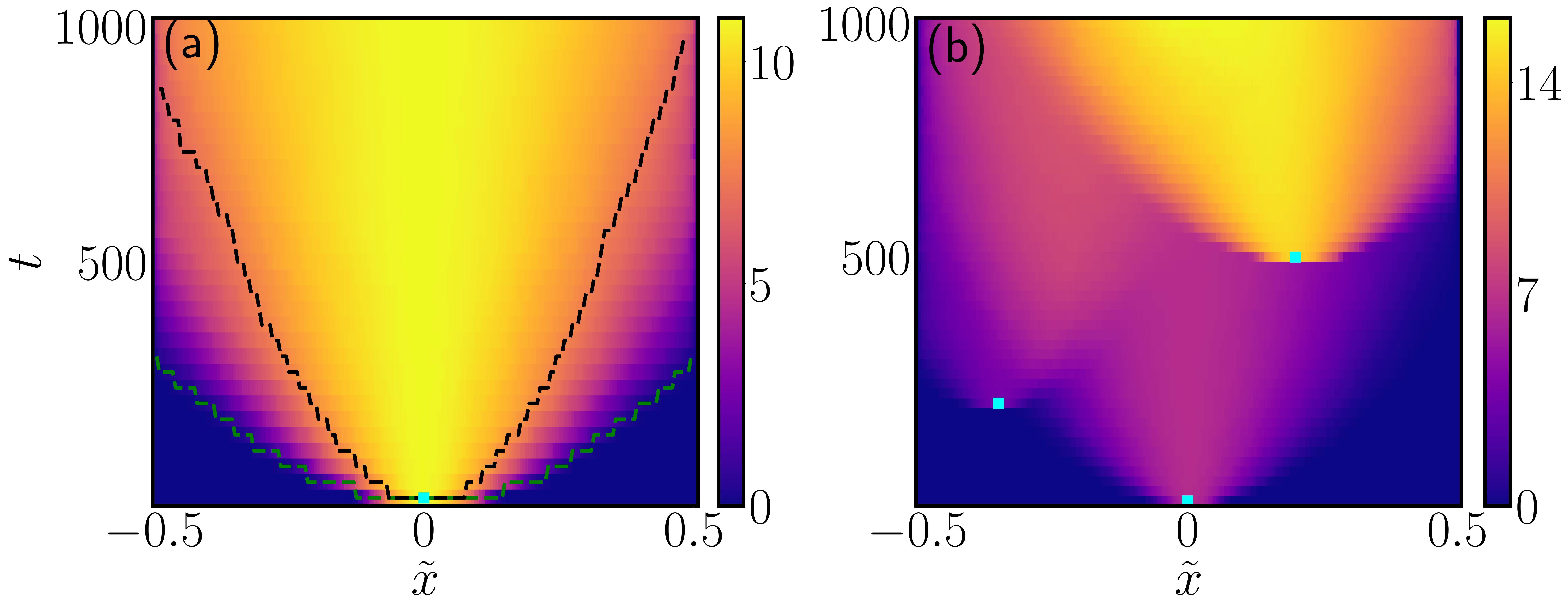}
    \caption{Heat map of ensemble-averaged entanglement entropy evolution showing the entanglement light cone. x-axis is the re-scaled spatial coordinate $\tilde{x}$ and y-axis is time $t$. The color indicates the entanglement entropy between the left and right part of the system, divided by the boundary at $\tilde{x}$. Number of modes $M=201$. (a) Single squeezer $r=8$ is placed at the center of the system (shown as a cyan box). The green dashed line represents the entanglement light cone ($t=T_1$) and the black dashed line represents the characteristic time-scale $T_1+T_2$ of entanglement-growth. See Sec.~\ref{sec:light_cone} for details. (b) Multiple squeezers placed at different spacetime (shown as cyan boxes). The first squeezer $r_1=5$ is placed at $\tilde{x}=0, t=0$, the second one $r_2=3$ is placed at $\tilde{x} = -0.35, t=200$ and the third one $r_3=7$ is placed at $\tilde{x} = 0.2, t=500$.
    \label{fig:lightcone}
    }
\end{figure}

\begin{figure}
\centering
\subfigure{\includegraphics[width = 0.45\textwidth]{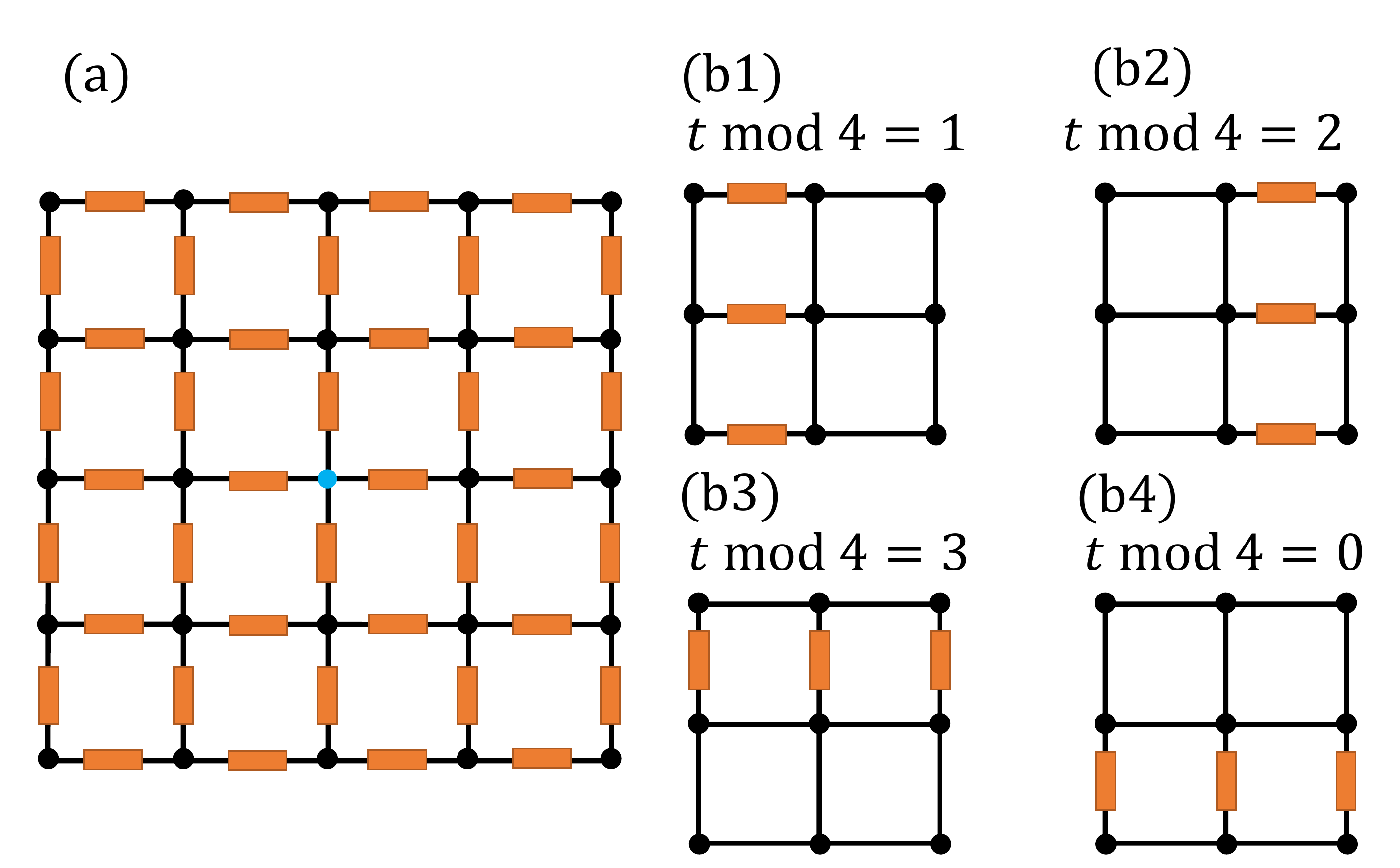}}
\subfigure{\includegraphics[width = 0.45\textwidth]{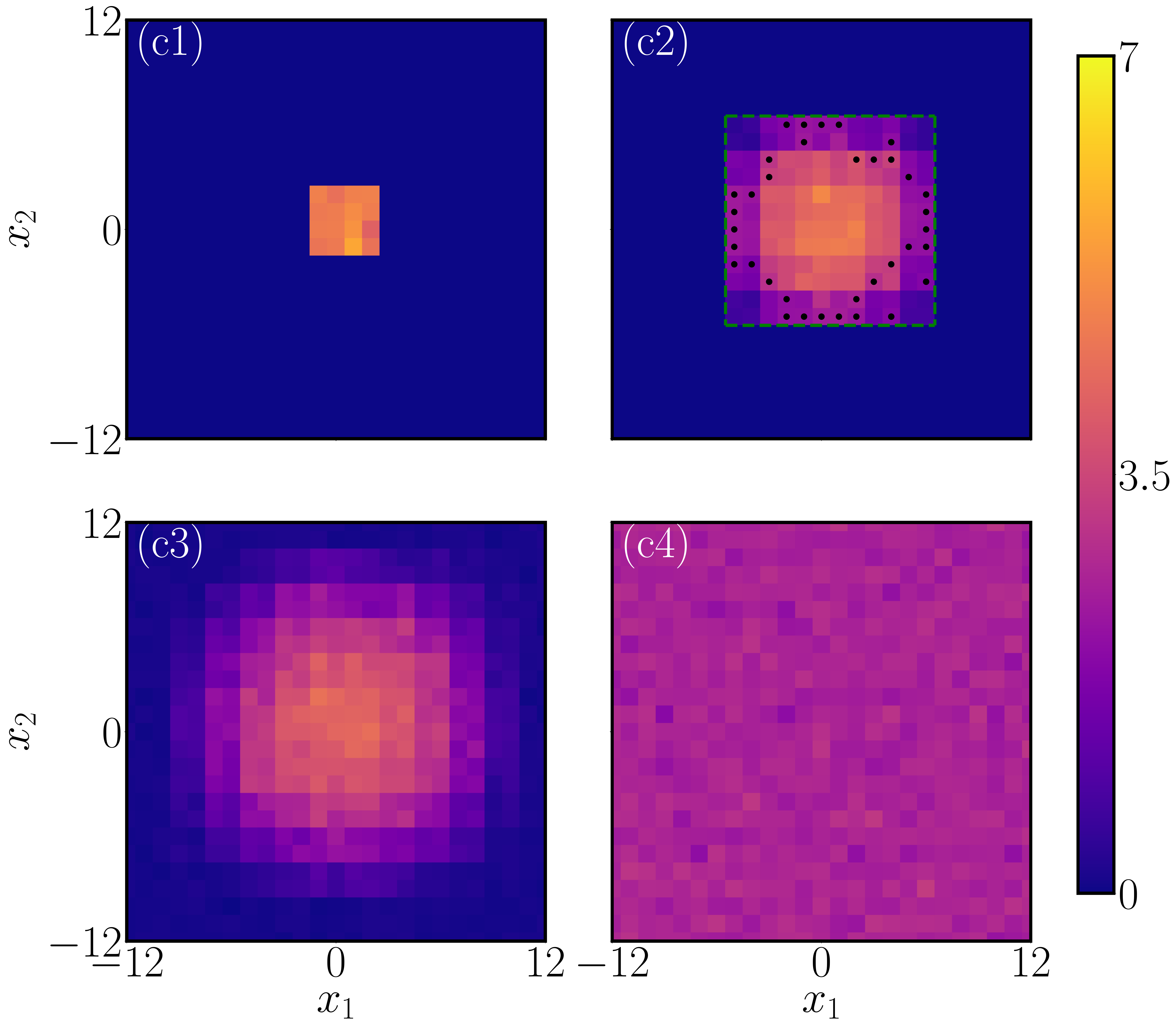}}
\caption{(a) Schematic graph of a 2-D random local circuit. Circles represent modes, and the cyan color indicating the location of a squeezer. 
(b1)-(b4) The random unitary gates are applied to system repeating four steps.
(c1)-(c2) The corresponding ensemble-averaged entanglement entropy in a 2-D system of $25\times 25$ modes. A single squeezer of $r=5$ is placed at the center of system at $t=0$. Subsystem $\calL$ is chosen to be each mode at $\bm x=(x_1,x_2)$. Results with alternative choices of $\calL$ can be found in Fig.~\ref{fig:2D_squares}. The snapshots are taken at $t=4, 12, 20, \infty$ from (c1) to (c4), and the green dashed and black dotted line in (c2) indicates the light cone ($t=T_1$) and the characteristic time-scale $T_1+T_2$. See Sec.~\ref{sec:light_cone} for details.
\label{fig:schematic_2D}
}
\end{figure}

Unitaries are applied on the edges $\calE$. We separate the edges into disjoint sets $\{\calE_{k}\}_{k=1}^K$, such that the edges in each set $\calE_k$ do not have common vertices. The dynamics repeat in a period of $K$ steps; in the $k$-th step of each period, one applies unitaries $U_{t,\bm x,\bm x^\prime}$ on each edge $\overline{\bm x\bm x^\prime}\in \calE_k$. The particular separation of the unitaries is not essential to the dynamics and equilibrium. As an example, in a 1-D local circuit, $K=2$ and we alternative between gates $\{U_{t,k,k+1}\}$ on $k$ odd and even; in a 2-D local circuit, we have $K=4$, as shown in Fig.~\ref{fig:schematic_2D}.

To produce the Gaussian states that enables various applications in communication, sensing and computing, we consider Gaussian unitaries~\cite{Weedbrook_2012}, which are unitaries generated by Hamiltonians that are second order in the quadrature operators (see Appendix~\ref{app:random_mat}). Gaussian unitaries include squeezing, which creates asymmetry in quadrature noises; and passive linear optics, which includes beamsplitters and phase-shifters.

Squeezing is essential for entanglement generation. However, as an `active' component, squeezing is relatively difficult to implement. Thus, we consider the gates $\{U_{t,\bm x,\bm x^\prime}\}$ to be passive linear-optics gates. And squeezing operations are added in between in a sparse way. As an example, in Fig.~\ref{fig:schematic_1D}(b), to establish entanglement, in this case a single vertex performs a squeezing operation (the cyan box), and then entanglement is generated by passing it around through passive components (the orange boxes), with vacuum on the other input modes. 

We expect random protocols to reveal universal characteristics, therefore we choose the passive linear optics gates $\{U_{t,\bm x,\bm x^\prime}\}$ to be Haar random (see Appendix~\ref{app:random_mat}). This is also justified by the following reasons: (1) In classical complex network theory~\cite{barabasi1999emergence,watts1998collective,barabasi1999mean}, various networks can be modeled as random networks with a proper degree distribution. (2) In condensed matter theory, random quantum circuits and Hamiltonian systems~\cite{nahum2017quantum,nahum2018operator,Keyserlingk:2018aa,Khemani:2018aa,Rakovszky:2018aa,you2018entanglement} are able to capture the essential quantum information spreading features in generic many-body interacting systems. (3) In real quantum networks, the form of entanglement required can be complicated, depending on the purpose, e.g. the weights of the global parameter of interest in a distributed sensing protocol~\cite{zhuang2018distributed}.

We aim to characterize the entanglement dynamics in the above random circuits. The entanglement entropy, measured by von Neumann entropy or Renyi entropy, can be numerically evaluated efficiently (see Appendix~\ref{app:entropy}). Two examples of time evolution of von Neumann entropy in 1-D are given in Fig.~\ref{fig:lightcone}. In Fig.~\ref{fig:lightcone}(a), we have a single squeezer at the center in the first step, which is identical to the case depicted in Fig.~\ref{fig:schematic_1D}. The entanglement entropy grows diffusively from the source of squeezing (see Sec.~\ref{sec:dynamic}). Different from the DV case, we can identify an entanglement light cone (green lines), which is the boundary between regions with substantial entanglement and regions with almost-zero entanglement. These phenomena can also be found in higher dimensional random local circuits, as shown in Fig.~\ref{fig:schematic_2D} for the 2-D case. When choosing a subsystem $\calL$ as an individual mode at $(x_1,x_2)$, we can see similar entanglement light cone (green dashed).

As shown in Fig.~\ref{fig:general_graph}(b), the above entanglement dynamics can be solved by mapping to a random walk on a graph, which gives the exact ensemble-averaged entanglement entropy (see Sec.~\ref{sec:mapping}) $\braket{S(\calL,t)}$ as a function of the total probability $\eta_{\calL,t}$ of having the walker in region $\calL$ in the corresponding random walk,
\be 
S(\eta_{\calL,t})
= g\left[\left(\sqrt{1+4\eta_{\calL,t}(1-\eta_{\calL,t})\sinh^2(r)}-1\right)/2\right],
\label{SL_preliminary}
\ee 
where 
$
g(x)  
$
is the (von Neumann or Renyi) entropy of a thermal state with mean photon number $x$ (see Appendix~\ref{app:entropy}) and $r$ is the original squeezing strength.
Note that the mapping holds for arbitrary graphs beyond the local Cartesian graphs shown above. The scrambling time---the time that the entire system becomes maximally entangled---can be calculated by the mixing time of the random walk (see Sec.~\ref{sec:mixing_time}). The mapping also gives the Page curves---the late-time equilibrium entanglement entropy
\be 
\braket{S(\calL,\infty)}= S({|\calL|}/{|\calG|})
\label{SL_page_preliminary}
\ee 
as the CV analog to Page curve (see Sec.~\ref{sec:equi_solu_general}), while the fluctuation can be solved as $\propto |\calL||\calR|$. Moreover, when there are multiple squeezers, we can regard the entanglement dynamics as the superposition of all single-squeezer dynamics, as depicted in Fig.~\ref{fig:lightcone}(b) and will be detailed in Sec.~\ref{sec:superposition}. When there are multiple squeezers, we find that the Page curve is determined by the average squeezing strength and density of the squeezers (see Sec.~\ref{sec:Page}). Therefore, combining the results, we have a complete understanding of the entanglement dynamics in a CV quantum network.

\section{Statistical theory of random CV quantum networks}
\label{sec:statistical_theory}

In this section, we present a statistical theory of the entanglement growth. We will focus on the single squeezer case in Fig.~\ref{fig:lightcone}(a), while the extension to multiple squeezers is presented in Sec.~\ref{sec:multiple_squeezers}. We introduce the mapping between random unitary circuits and random walk on a graph in Sec.~\ref{sec:mapping}, which allows us to solve the Page curves in Sec.~\ref{sec:equi_solu_general} and scrambling time in Sec.~\ref{sec:mixing_time} for general graphs. Explicit closed-form solutions can be obtained for local Cartesian graphs of an arbitrary dimension in Sec.~\ref{sec:dynamic}. Finally, we present the entanglement witness for multipartite entanglement in Sec.~\ref{sec:multipartite_E}.

\subsection{Mapping to random walk on graphs}
\label{sec:mapping}
Consider the entire unitary evolution $U(t)$ of the random circuit. In the single-squeezer case, the mode annihilation operator $a_{\bm x,t}$ at vertex $\bm x\in \calG$ experiences a passive transform, which in general can be expresses as 
\be 
a_{\bm x,t}=e^{i\theta_{\bm x,t}}\sqrt{w_{\bm x,t}} a_{\rm SV}+{\rm vac},
\label{axt}
\ee 
where mode $a_{\rm SV}$ is in a squeezed-vacuum (SV) state with strength $r$ and `vac' denotes all vacuum terms that complete the commutation relation. Here the phase $\theta_{\bm x,t}$ is entirely random, and the positive weights $\bm w_t\equiv \{w_{\bm x,t}\}_{\bm x\in \calG}$ describe the overall energy splitting of the single SV among all modes.

For any subsystem $\calL$ with a density operator $\rho_\calL(t)$, we can design a passive linear optics circuit $U_{\calL,t}$ such that $U_{\calL,t}\rho_\calL(t)U_{\calL,t}^\dagger$ concentrates all the squeezing parts to a single mode
\be
a_{\calL,t}=\sqrt{\eta_{\calL,t}}a_{\rm SV}+{\rm vac},
\ee 
with the total transmissivity
\be 
\eta_{\calL,t}=\sum_{\bm x\in\calL}w_{\bm x,t},
\label{etat_def}
\ee 
and all other modes are in vacuum required by energy conservation. Because unitary operations preserve entropy, the entanglement entropy of $\calL$ can be calculated from the entropy of mode $a_{\calL,t}$ as
\begin{align}
&S\left(\calL,t\right)=S\left(\eta_{\calL,t}\right)
\label{SL}
\\
&\simeq \frac{1}{2}\log_2\left[\eta_{\calL,t}\left(1-\eta_{\calL,t}\right)\right]+\frac{1}{\ln2}\left(r+1\right)-1,
\label{SL_sim}
\end{align}
where $S(\eta_{\calL,t})$ is defined in Eq.~(\ref{SL_preliminary}).
We will focus on von Neumann entropy, but all of our results can be adapted to Renyi entropy easily.
At the large squeezing limit of $\sqrt{\eta_{\calL,t}(1-\eta_{\calL,t})}e^r\gg1$, for von Neumann entropy we have Eq.~(\ref{SL_sim}) to the leading order. When $\eta_{\calL,t}=0,1$, subsystem $\calL$ has zero or entire portion of the SV, indeed from Eq.~(\ref{SL}) we have $S\left(\calL,t\right)=0$. When $\eta_{\calL,t}=1/2$, we have the maximum entropy $S_0(r)=g(\sinh^2(r/2))\simeq \log_2(e^r/4)$. When one has large number of modes, this should agree with the result of $\max_{\calL} S\left(\calL,t\right)$ at any time.

So far we have the {\it exact} result $S\left(\eta_{\calL,t}\right)$ of the entanglement entropy of an arbitrary subsystem $\calL$, given the weights $\bm w_t$ (which determines $\eta_{\calL,t}$) obtained in each random circuit realization. 
Due to the self-averaging in the random circuit, we expect 
\be 
\braket{S\left(\eta_{\calL,t}\right)}=S\left(\braket{\eta_{\calL,t}}\right)
\ee 
up to corrections that decay with the system size. Thus, we have reduced the problem of solving the ensemble-averaged entanglement dynamics to solving the ensemble-averaged dynamics of the weights $\braket{\bm w_t}$. 
\begin{figure}
    \centering
    \subfigure{\includegraphics[width = 0.45\textwidth]{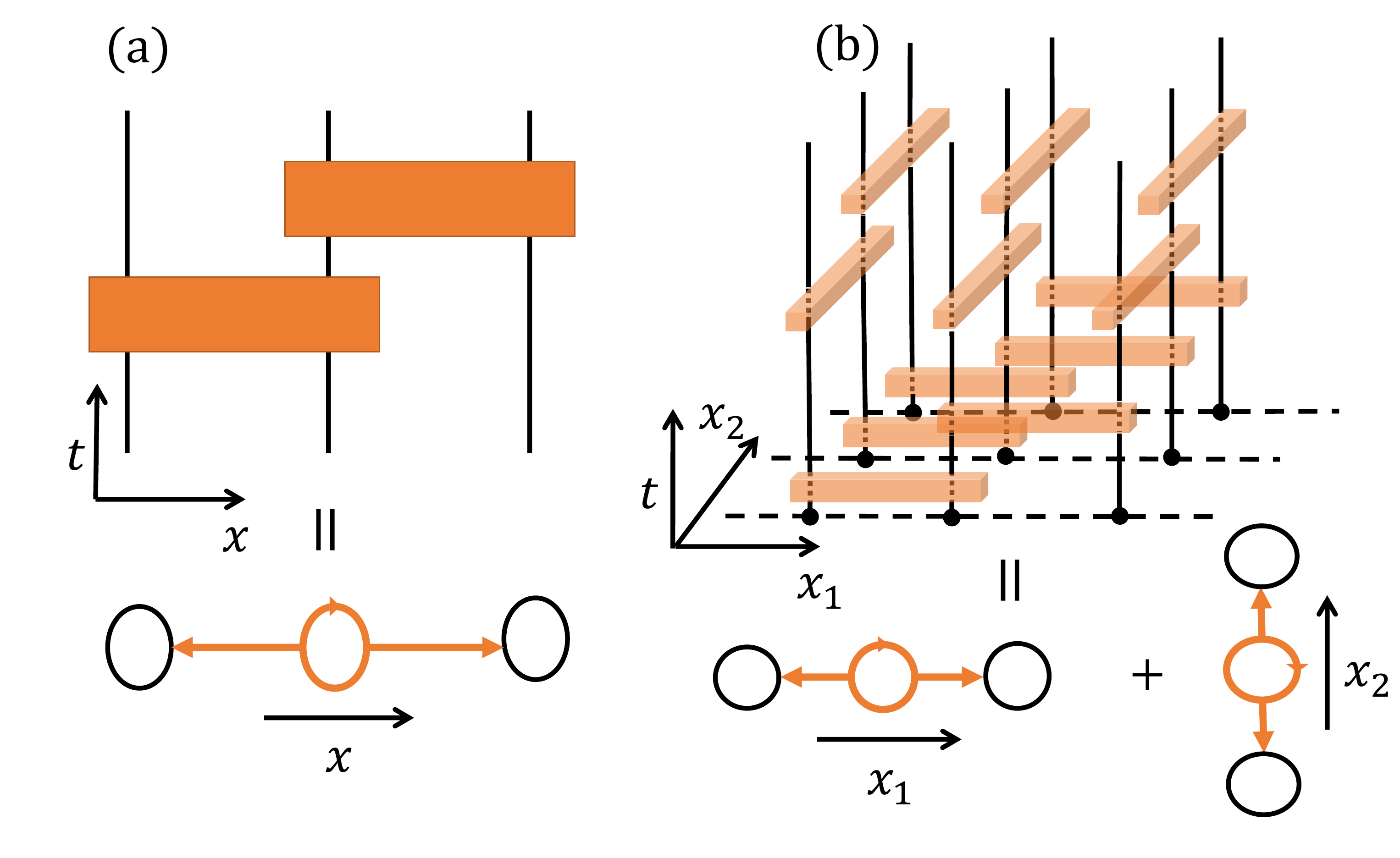}}
    \subfigure{\includegraphics[width = 0.45\textwidth]{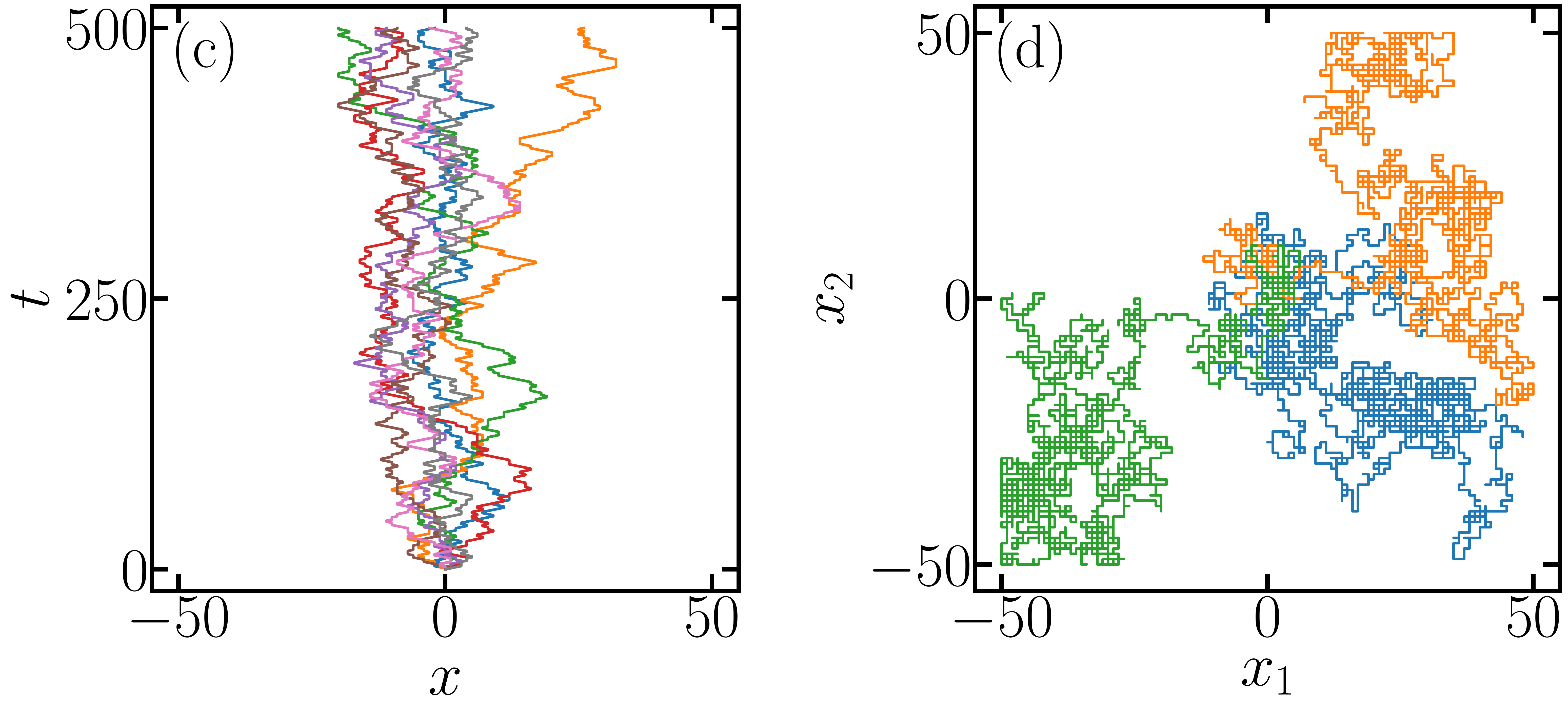}}
    \caption{Random circuits and the corresponding modified random walk in 1-D and 2-D Cartesian graphs $\calG_D$. (a), (b) Schematic of a $K$-step cycle of gates that implement a random-walk step. The random walk along each dimension is independent, as shown in (b). (c), (d) Examples of the modified random walk in 1-D and 2-D, following the rule in Eq.~(\ref{total_transition_matrix}). In 1-D, the time trajectories up to $t=500$ of different instances are plotted in different colors. In 2-D, the trajectories are shown up to $t=5000$ for three different instances. 
    \label{ab font larger}
    \label{fig:random_walk}
    }
\end{figure}

We start by focusing on a single gate $U_{t,\bm x,\bm x^\prime}$ on the modes at $\bm x$ and $\bm x^\prime$.
By considering the Haar random ensemble averaging, we can derive the exact equation of motion of the weights as (see Appendix~\ref{app:random_derivation})
\be 
\braket{w_{\bm x,t+1}}=\braket{w_{\bm x^\prime,t+1}}=\frac{1}{2}\left(\braket{w_{\bm x,t}}+\braket{w_{\bm x^\prime,t}}\right).
\label{ensemble_dynmaics}
\ee 
The overall dynamics alternatives in $K$ steps, in the $k$-th step the transition of Eq.~(\ref{ensemble_dynmaics}) on all edges in $\calE_k$ is applied. 

An immediate observation from Eq.~\ref{ensemble_dynmaics} is that the change of the entanglement entropy of $\calL$, determined by the total weights $\braket{\eta_{\calL,t}}$, is related only to boundary $\partial \calL$ (schematic in Fig.~\ref{fig:general_graph}(c)), in the sense that
\begin{align} 
&\braket{\eta_{\calL,t+1}}-\braket{\eta_{\calL,t}}=
\nonumber
\\
&\frac{1}{2}\left[\sum_{\bm x \in \partial \calL^+} \braket{w_{\bm x,t}}- \sum_{\bm x \in \partial \calL^-} \braket{w_{\bm x,t}}\right],
\label{eta_boundary}
\end{align}
which equals the net flow of the weights from the vertices on the outer boundary $\partial \calL^+$ towards $\calL$ and the weights from the inner boundary $\partial \calL^-$ out from $\calL$.

Another observation is that the weights update rule in Eq.~(\ref{ensemble_dynmaics}) also describes the probability evolution of a lazy symmetric random-walk step, where the walker have half probability of staying and half probability of taking a step along $\overline{\bm x\bm x^\prime}$ (see Fig.~\ref{fig:random_walk}). Combining the $K$ steps, the underlying transition matrix for the weights
\be
\mathbb{E}_{\bm x, \bm x^\prime} = \prod_{k = 1}^{K} \frac{1}{2}\left(I+E_{k,\bm x,\bm x^\prime}\right),
\label{total_transition_matrix}
\ee
where $I$ is the identity matrix and $E_{k,\bm x,\bm x^\prime}$ describes the adjacency matrix for the corresponding graph $(\calG,\calE_k)$ (an isolated mode is regarded as a vertex with a loop). Eq.~\ref{total_transition_matrix} describes a modified symmetric random walker on the graph (see Fig.~\ref{fig:random_walk}), with $K$ steps combined to implement a single random-walk step from the current position $\bm x$ to all neighbors $\calN(\bm x)$ (including $\bm x$) with equal probability.

Utilizing the $K$-step transition matrix, the ensemble-averaged weights can be solved at any time $t$ as
\be
\braket{\bm w_t} = \braket{\bm w_0}\mathbb{E}_{\bm x, \bm x^\prime}^{[t/K]},
\label{ensemble_dynamics_matrix}
\ee
with the initial condition $\braket{\bm w_0}=\delta_{\bm x_0}$ as the Kronecker delta at the squeezer position $\bm x_0$. Thus, one can obtain $\braket{\eta_{\calL,t}}$ and the exact result of $S(\braket{\eta_{\calL,t}})$ from Eq.~(\ref{SL}) on any graph.

\begin{figure}
    \centering    
    \includegraphics[width = 0.45\textwidth]{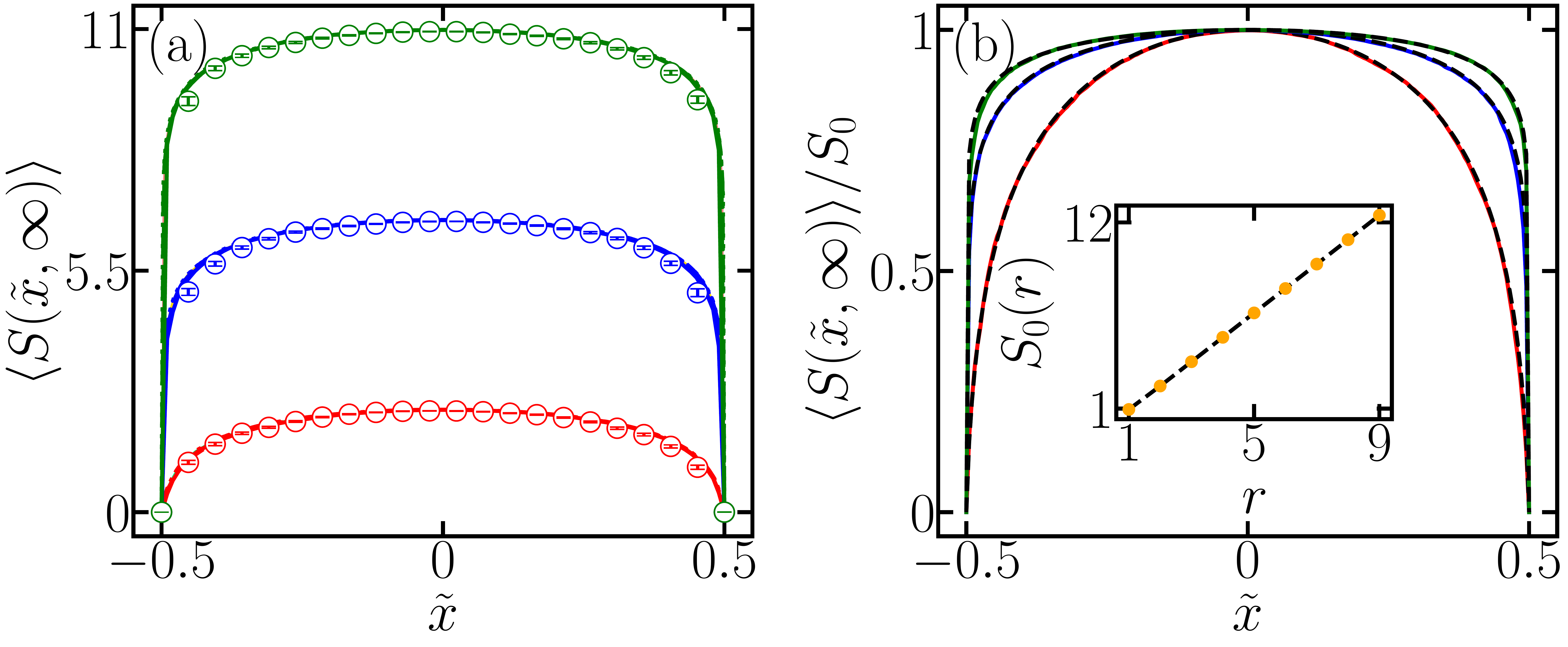} 
    \caption{
    (a) 
    Page curves with a single squeezer, $r=2, 5, 8$ shown by the red, blue and green lines separately. The overlapping scatter points, solid, dashed, dash-dot and dotted lines corresponds to system of $M=21, 101, 201, 301, 401$. Error bars for the case $M=21$ lie in the empty circles. Invisible shadow area shows the numerical precision.  
    (b)
    Re-scaled Page curves in (a). Black dashed lines show the theory results of Eq.~(\ref{SL_page}). The inset is the dependence of maximum height $S_0$ on the single squeezer $r$. Numerical results in system $M=201$ (orange dots) and analytical results (black dashed line) Eq.~(\ref{SL}) with $\eta_{\calL,t}=1/2$ agree well. 
    \label{fig:pages_1D}
    }
\end{figure}

We give examples of the random walk in Fig.~\ref{fig:random_walk} in 1-D and 2-D Cartesian graphs, whose entanglement evolution can be found in Figs.~\ref{fig:schematic_1D} and~\ref{fig:schematic_2D}. For the later use, we also introduce a general $D$ dimensional Cartesian lattice $\calG_D$, with the coordinates $\bm x=(x_1,
\cdots, x_D)$ on a grid ($x_d\in[-N,N]$). The total number of modes is $|\calG_D|=M^D$, with $M=2N+1$ modes on each dimension.

In the following, we will consider the equilibrium and the dynamics. Some results hold for general graphs, while some analytical results are made possible by considering the special case of $\calG_D$.

\subsection{Equilibrium of CV random networks: Page curves and fluctuations}
\label{sec:equi_solu_general}

In this section, we focus on the Page curves---the equilibrium entanglement distribution at infinite time. In order to share entanglement, squeezers are applied, which are then followed up by the random beamsplitters and phase shifters. As the layers of gates increases, the overall passive linear transform will approach the Haar measure (see Appendix~\ref{app:random_mat}). Therefore, we can regard the equilibrium entanglement distribution as the CV analog to Page curves.

\begin{figure}
    \centering
    \includegraphics[width = 0.45\textwidth]{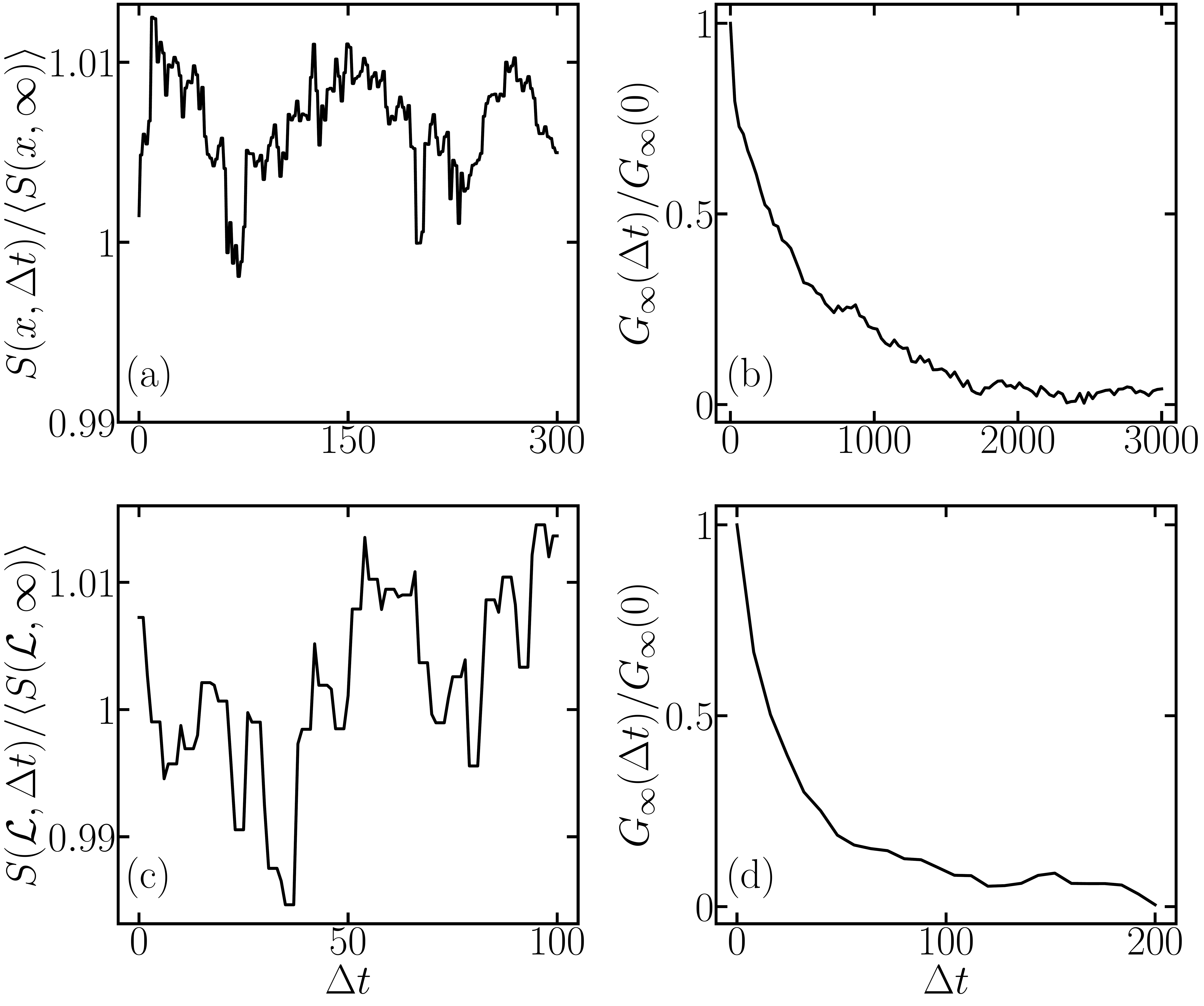} 
    \caption{Entanglement entropy evolution for a period of $\Delta t$ at late time $t=\infty$ (left panel, a and c) and normalized auto-correlation $G_\infty(\Delta t)/G_\infty(0)$ (right panel, b and d) in a 1-D Cartesian system of $M=61$ modes (first row, a and b) and 2-D Cartesian system of $11\times11$ modes (second row, c and d, ) correspondingly. A squeezer $r = 5$ is placed at the center of system. The subsystem $\calL=[-30, -10]$ contains modes to the left of $x=-10$ in the 1-D system, and $\calL = [-5, 0]\times[-5, 0]$ contains a quarter of the square in the 2-D system. The auto-correlation decays significantly at the order of mixing time.
    \label{fig:auto_correlation}
    }
\end{figure}

Considering the mapping from the circuit to the random walk, the equilibration of the entanglement also corresponds to the full mixing of the random walk on the graph.
Assuming the full connectivity of the graph, due to the special transform matrix in Eq.~(\ref{total_transition_matrix}), the equilibrium (stationary) state of weights is uniform among all vertices, i.e.,
\be 
\braket{\bm w_{\bm x,\infty}}=1/|\calG|,
\label{w_stationary}
\ee 
where $|\calG|$ is the total number of vertices, despite how one arranges the set of edges $\calE_k$. Note that this is different from normal random walks on a graph, where the stationary state has weights proportional to the degree of the vertex $|\calN(\bm x)|$~\cite{lovasz1993random}. Therefore, the total transmissivity (i.e., total weights)
\be 
\braket{\eta_{\calL,\infty}}={|\calL|}/{|\calG|}.
\ee 
In fact, assuming fully random weights from a Haar random unitary, one can obtain the probability density of total weights as (see Appendix~\ref{app:eta_var})
\be 
P(\eta_{\calL,\infty}=\eta)\propto\eta^{|\calL|-1}(1-\eta)^{|\calR|-1}.
\label{P_eta_dis}
\ee 
From Eq.~(\ref{SL}), the Page curve is therefore given by
\begin{align}
\braket{S(\calL,\infty)}&= S(\frac{|\calL|}{|\calG|})
\label{SL_page}
\\
&\simeq \frac{1}{2}\log_2\left[\frac{|\calL|}{|\calG|}\left(1-\frac{|\calL|}{|\calG|}\right)\right]+\frac{1}{\ln2}\left(r+1\right)-1.
\label{SL_page_sim}
\end{align}
The maximum is achieved at ${|\calL|}/{|\calG|}=1/2$, which equals $\braket{S_0(r)}$ introduced following Eq.~(\ref{SL}). The second equality is the leading order result similar to Eq.~(\ref{SL_sim}).

Note that in terms of Page curves, the graph topology is irrelevant as the entire dynamics is equivalent to a single passive global Haar unitary; therefore, we can simply stretch the coordinates $\bm x$ of a general graph $\calG$ to a single coordinate $x$ in 1-D.
It then suffices to verify our theory of Page curves in $\calG_1$, which allows simple visualization. In the 1-D system, when subsystem $\calL$ contains the left side of mode $x$, we have 
$
\braket{\eta_{\calL,\infty}}={|\calL|}/{|\calG|}=({x+N})/({2N+1})\simeq\tilde{x}+1/2.
$ 
Therefore, it is convenient to choose the parameterization $\tilde{x}$.
Fig.~\ref{fig:pages_1D}(a) plots $\braket{S(\tilde{x},\infty)}$ for various squeezing values of $r$ and system sizes of $M$, where we see perfect overlapping among curves with identical $r$ for different system sizes. And they all agree with Eq.~(\ref{SL_page}) very well, as shown in Fig.~\ref{fig:pages_1D} (b). As a by-product, the maximum entanglement---the maximum height of the Page curve $\max_x \braket{S(x,\infty)}$---agrees with the theory prediction $\braket{S_0(r)}$ following Eq.~(\ref{SL}), as shown in Fig.~\ref{fig:pages_1D}(b) inset.

\begin{figure}
    \centering
    \subfigure{\includegraphics[width = 0.45\textwidth]{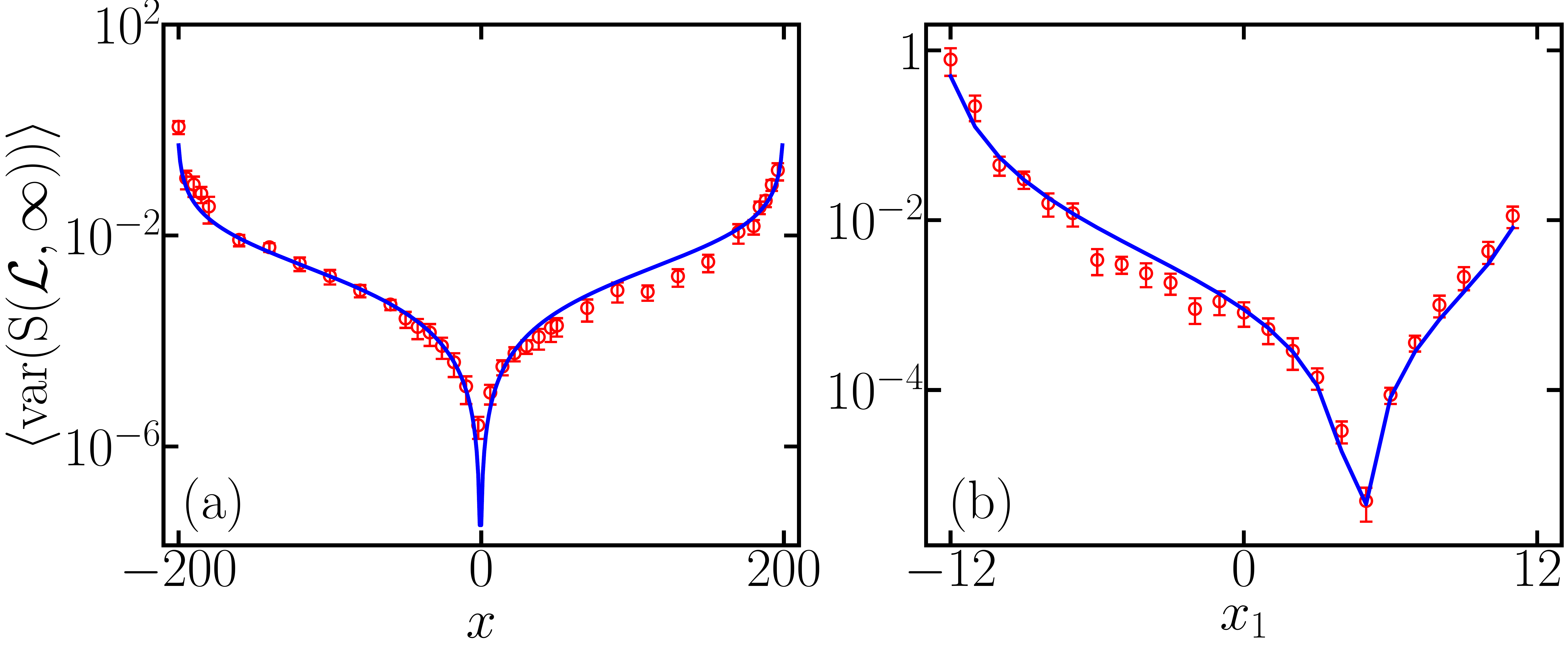}}
    \caption{Variance of Page curves, numerical results (red dots with error bars) agree well with Eq.~(\ref{eq:fluctuation_static}) (blue curves), in
    (a) a 1-D system of $|\calG|=401$ with a squeezer $r = 8$. 
    (b) a $25\times 25$ 2-D system with a squeezer $r=5$.
    Subsystem $\calL$ is chosen to be (a) a line $\calL=[-200,x]$ and (b) a square $\calL=[-12,x_1]\times [-12,x_1]$.
    \label{fig:static_fluctuation}
    }
\end{figure}

Furthermore, we can also consider the fluctuations around the Page curves, as exemplified in Fig.~\ref{fig:auto_correlation} (a) and (c). Since the entanglement entropy is determined by $\eta_{\calL,t}=\sum_{\bm x\in \calL}w_{\bm x,t}$, it takes some time for $\{w_{\bm x,t}\}_{\bm x\in \calL}$ to entirely change its values; thus, there will be correlations in the entanglement entropy at different times. We consider the equilibrium auto-correlation 
\begin{align} 
&G_\infty\left(\Delta t\right)=\lim_{t\to\infty}
\nonumber
\\
&\braket{(S(\calL,t)-\braket{S(\calL,\infty)})(S(\calL,t+\Delta t)-\braket{S(\calL,\infty)})}.
\end{align}
We expect the decay of this auto-correlation should have a similar time-scale with the mixing time of the entire system (which will be detailed in Sec.~\ref{sec:mixing_time}), as demonstrated in Fig.~\ref{fig:auto_correlation}. 
When $\Delta t=0$, the auto-correlation goes to the variance
\begin{align} 
&G_\infty\left(0\right)= \braket{{\rm var}\left(S(\calL,\infty)\right)},
\\
&= 
 \left(\left.\frac{\partial S\left(\eta_{\calL,t}\right)}{\partial \eta_{\calL,t}}\right\vert_{\eta_{\calL,t}=\frac{|\calL|}{|\calG|}}\right)^2 \frac{|\calL||\calR|}{|\calG|^2(|\calG|+1)},
\label{eq:fluctuation_static}
\end{align}
where we have used the chain rule of variance and the variance of $\eta_{\calL,\infty}$ can be obtained in Appendix~\ref{app:eta_var}. Numerical results in 1-D and 2-D Cartesian graphs $\calG_D$ agree well with Eq.~(\ref{eq:fluctuation_static}), as shown in Fig.~\ref{fig:static_fluctuation}.

On the other hand, if we look at the change of subsystem entropy within a short period of time, e.g., a single-step, similar to Eq.~(\ref{eta_boundary}) the short-time fluctuating $\braket{\left(S\left(\calL,t+1\right)-S\left(\calL,t\right)\right)^2}$ will mainly come from the boundary. This interplay of short time fluctuation related to the boundary, while long time fluctuation related to the bulk manifests the rich entanglement dynamics in random quantum networks.

\subsection{Entanglement scrambling time}
\label{sec:mixing_time}

\begin{figure}
    \centering
    \includegraphics[width = 0.45\textwidth]{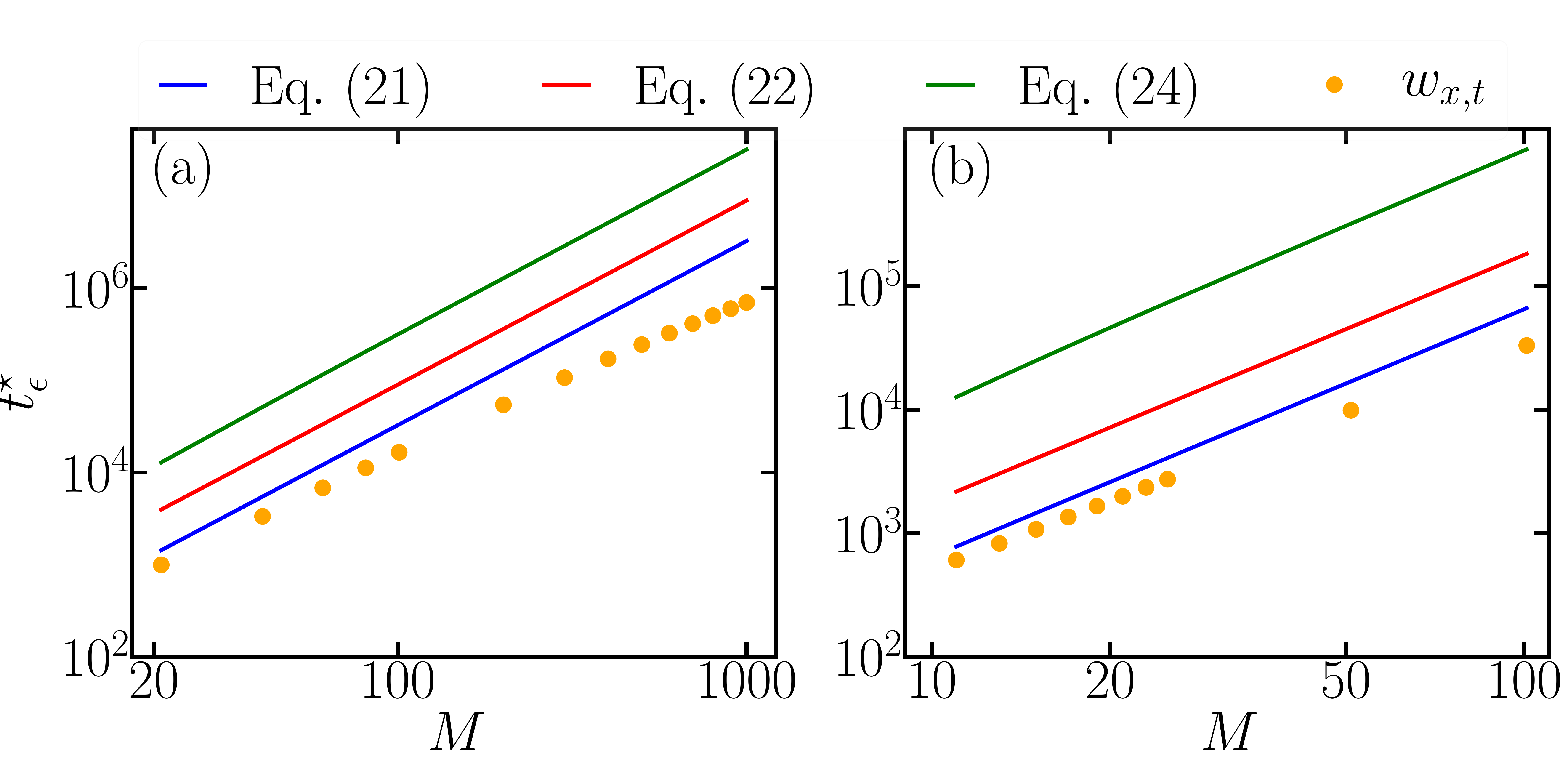}
    \caption{Three estimators for the mixing time $t_\epsilon^\star$ in 1-D system and 2-D grid system of different number of nodes $|\calG|=M^2$, shown in (a) and (b). Blue, red and green lines show the bounds for mixing time by Eq.~(\ref{graph_bound_1}), ~(\ref{graph_bound_2}) and ~(\ref{graph_bound_3}). Orange scatter dots show the numerical mixing time for corresponding random walk. $\epsilon$ is set to be $1\times10^{-7}$. Note: In 2-D grid system, we estimate the conductance without full optimization.
    \label{fig:mixing_time}
    }
\end{figure}

The entanglement scrambling time---the time for the system to be maximally entangled and reach the equilibrium is an important quantity for many physical problems, especially those related to the black hole~\cite{hayden2007black,sekino2008fast}. In the CV quantum network, this scrambling time can be directly obtained from the mixing time $t_\epsilon^\star$ of random walks on a graph, where the probability measure (weights) get $\epsilon$-close to the stationary state in Eq.~(\ref{w_stationary}). Formally, we define the mixing time $t_\epsilon^\star$ to be the time when the deviation $|\braket{w_{\bm x,t}}-{1}/{|\calG|}|\le \epsilon$ for all $\bm x\in \calG$. Multiple estimates can be obtained, as we explain bellow~\cite{lovasz1993random}.

The first estimate relies on the eigenvalues of the transition matrix $\mathbb{E}_{\bm x, \bm x^\prime}$ in Eq.~(\ref{total_transition_matrix}). The stationary state $\braket{w_{\bm x,\infty}}={1}/{|\calG|}$ corresponds to the largest eigenvalue of unity, and the second largest eigenvalue  $\lambda^\star<1$ gives the decay of deviations
\be 
|\braket{w_{\bm x,t}}-\frac{1}{|\calG|}|\sim {\lambda^\star}^{(t/K)},
\label{graph_bound_1}
\ee 
From the above, one can obtain $t_\epsilon^\star\sim K \ln \left(1/\epsilon\right)/\ln (1/\lambda^\star)$. Here we have taken into account that, one needs $K$ steps to implement the transition in Eq.~(\ref{total_transition_matrix}).

For the Cartesian graphs $\calG_D$, we can obtain the convergence towards $\braket{\bm w_{\bm x,\infty}}=1/|\calG|$ from the expected hitting time in the large $t$ limit as~\cite{lovasz1993random}
\be 
|\braket{w_{\bm x,t}}-\frac{1}{|\calG|}|\sim 6^{-t/DM^2},
\label{graph_bound_2}
\ee
where $M=|\calG|^{1/D}$ is the length of the Cartesian graph in each dimension.
This gives another estimate $t_\epsilon^\star\sim\ln\left(1/\epsilon\right)DM^2/\ln(6)$.

\begin{figure}
    \centering
    \includegraphics[width = 0.45\textwidth]{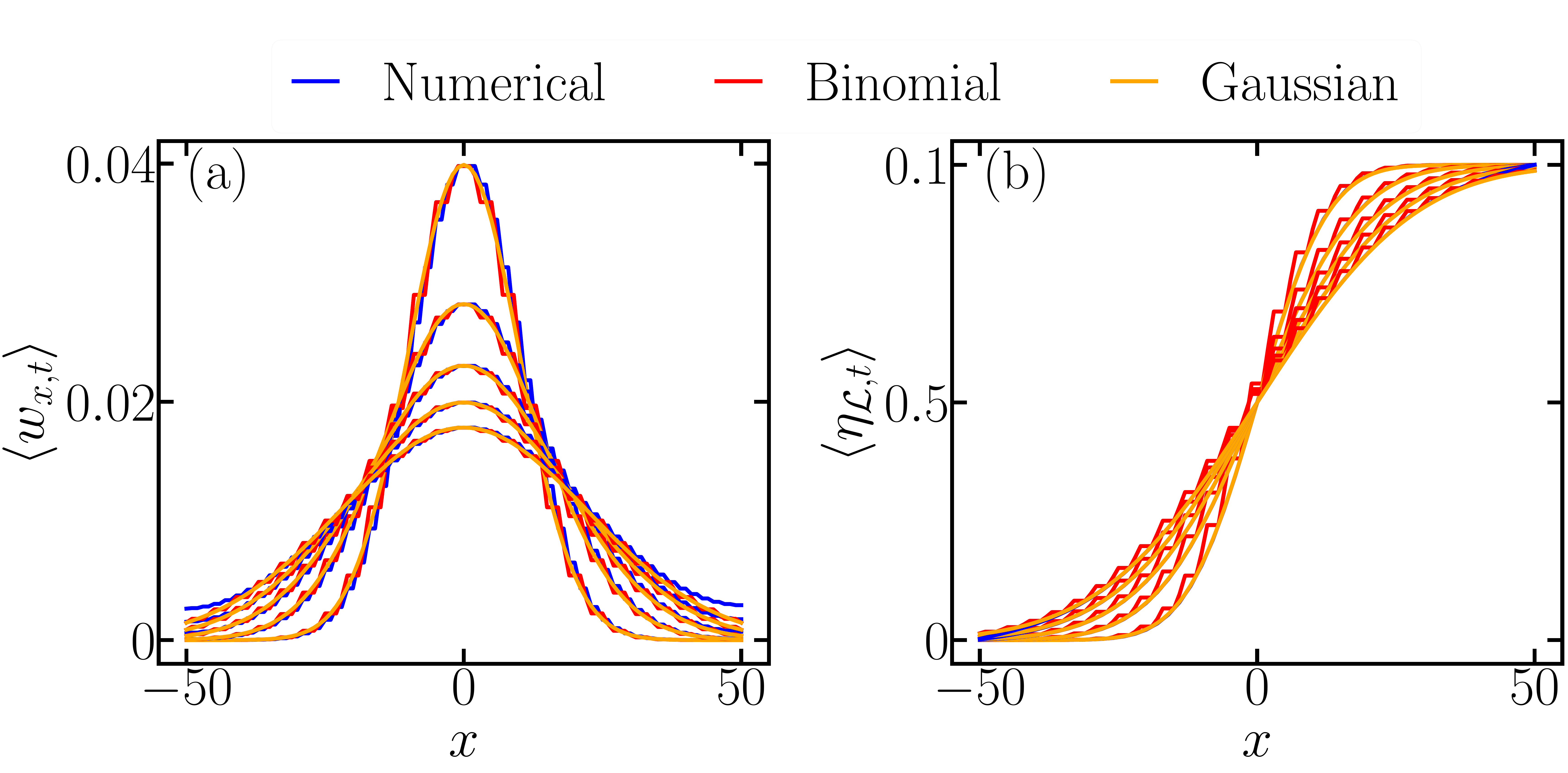}
    \caption{Comparison of solutions to weight $\braket{w_{\bm x,t}}$ and $\braket{\eta_{\calL,t}}$ in a 1-D system of $|\calG| = 101$ modes, shown in (a) and (b). (a) Curves from top to bottom show snapshots of weight field at $t=100, 200, 300, 400, 500$. (b) Curves from bottom to top (left part) show snapshots of $\braket{\eta_{ \calL,t}}$ field at $t=100, 200, 300, 400, 500$.
    \label{fig:1D_solution_compare}
    }
\end{figure}

A third bound can be obtained by calculating the conductance of a graph as
\be
\Phi(\calG)=\min_{\calL}\frac{|\nabla \calL||\calG|^2}{2|\calE||\calL||\calR|},
\label{conductance}
\ee
where $|\nabla \calL|$ denotes the set of boundary edges connecting $\calL$ to $\calR$ and $|\calE|$ is the total number of edges. The minimization will be able to capture the slowest part of the mixing. We can obtain from Ref.~\cite{lovasz1993random} as
\be
|\braket{w_{\bm x,t}}-\frac{1}{|\calG|}|\sim \left(1-\frac{\Phi^2}{8}\right)^t,
\label{graph_bound_3}
\ee
therefore the another estimate on the mixing time can be obtained as $t_\epsilon^\star\sim  \ln \left(\epsilon\right)/\ln \left(1-\Phi^2/8\right)$. 

For the Cartesian graph $\calG_D$ of length $M$ on each direction, the total number of edges of $\calG_D$ is
$
|\calE| = DM^{D-1}(M-1),
$
and total number of vertices in subsystem $|\calG_D| = M^D$. And we need $K=2D$ steps to implement a single symmetric walker step on each direction. For $D=1$, one has $|\nabla \calL|=1$, $|\calG_D|=M$ and $|\calE|=M-1$, therefore it is straightforward to obtain $\Phi(\calG_1)=2/(M-1)$. For $D=2$, we give an estimation from simple heuristic numerical minimization.

We compare the three estimates of mixing time on $\calG_D$ for $D=1,2$ in Fig.~\ref{fig:mixing_time}, and find a good agreement in the scaling of the scrambling time as $t_\epsilon^\star\propto D M^2$.

\subsection{Closed form solutions to Cartesian graphs}
\label{sec:dynamic}

\begin{figure}
    \centering
    \includegraphics[width = 0.45\textwidth]{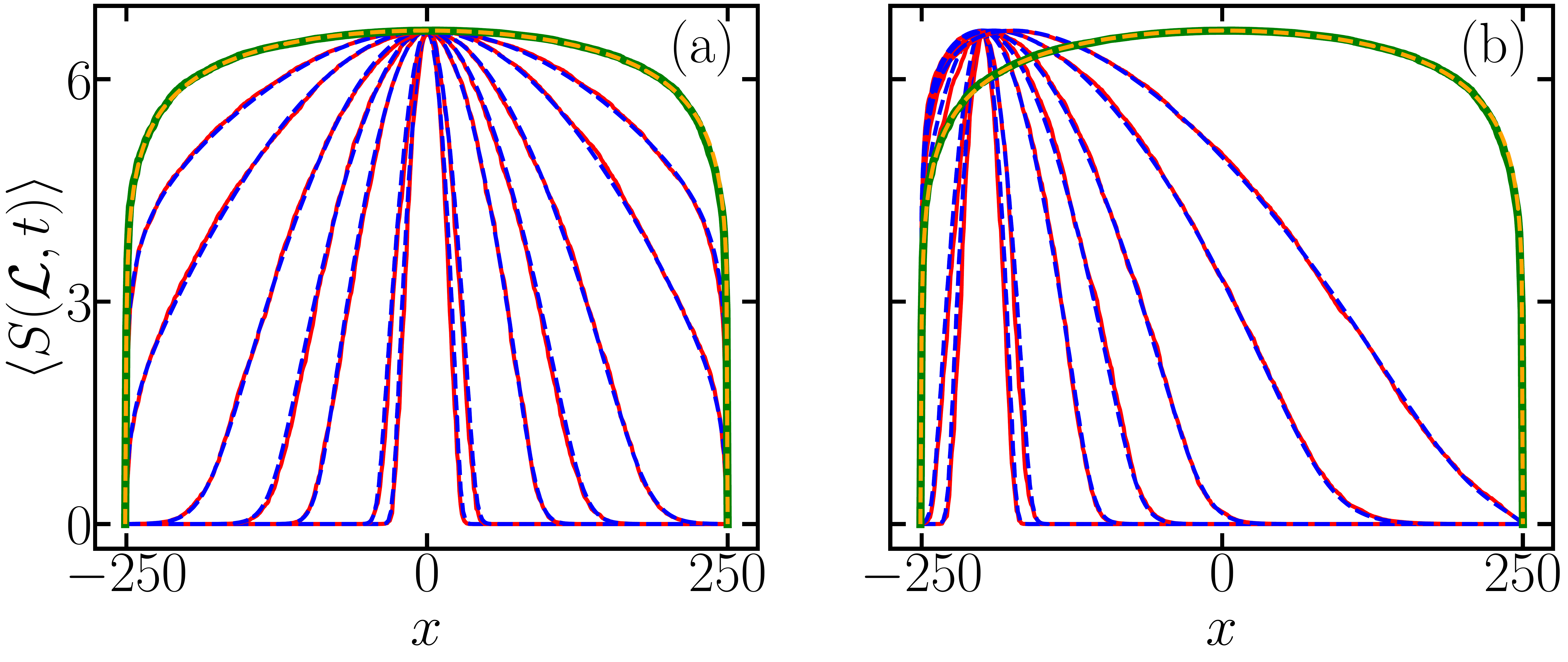}
    \caption{Ensemble-averaged entanglement dynamics in 1-D system of $|\calG|=501$ by weight field solution (red solid, obtained from Eq.~(\ref{SL}) and numerical solving Eq.~(\ref{ensemble_dynamics_matrix})) and real random circuit evolution (blue dashed). Curves from inside to outside corresponds to time at $t=50, 100, 500, 1000, 2000, 5000, 10000$. Green solid and orange dashed lines represent real Page curve and theoretical results in Eq.~(\ref{SL_page}). 
    (a) A single squeezer $r=5$ is placed at the center of the system $x=0$. 
    (b) The squeezer $r=5$ is placed at $x = -200$.
    \label{fig:1D_entropy_compare}
    }
\end{figure}

With the understanding of the equilibrium Page curves, we now proceed to characterize the dynamical evolution towards the equilibrium. We will focus on the Cartesian graphs $\calG_D$, which allows closed-from solutions by analog to random walkers. In the continuum limit, the dynamics can be well-described by a diffusive PDE (Sec.~\ref{sec:continuum}). In particular, we identify a unique parabolic entanglement light cone, followed by an entanglement sudden growth phenomenon, in CV networks (Sec.~\ref{sec:light_cone}), as has already been shown in Figs.~\ref{fig:lightcone} and \ref{fig:schematic_2D}.

In Cartesian graphs, the random walk analog to Eq.~(\ref{ensemble_dynmaics}) can be understood as independent along each dimension. Thus, we modify the Pascal's triangle from a usual random walk to obtain the solution
\be 
\braket{\bm w_t}^{(\rm Bi)}=\Big\{\frac{1}{2^{t+D}}\prod_{d=1}^D
{{n_t}\choose{n_{x_d,t}}}
\Big\}_{x_d=-N}^N,
\label{D_binomial}
\ee 
with $n_t=[t/D], n_{x_d,t}=[\frac{x_d}{2}]+[\frac{t}{2D}]$ and ${a\choose b}$ as the binomial factor of $a$-choose-$b$.

With the weights in hand, we can calculate the entanglement entropy of an arbitrary subsystem $\calL$.
For example, we can consider $\calL=\{\bm x^\prime| x_d^\prime<x_d, 1\le d \le D\}$, i.e., the system is cut into two parts by a high-dimensional plane. Then we can obtain the ensemble averaged total transmissivity
\begin{align}
&\braket{\eta_{\calL,t}}^{(\rm Bi)}=\sum_{\bm x\in \calL}\braket{ w_{\bm x^\prime,t}}^{(\rm Bi)}=\frac{1}{2^{t+D}}\prod_{d=1}^D2\sum_{n^\prime=0}^{n_{x_d,t}}  {n_t \choose n^\prime}=\prod_{d=1}^D
\nonumber
\\
&  \left[ 1-\frac{1}{2^t}{{n_t}\choose{1+n_{x_d,t}}}F(1,1-n_t+n_{x_d,t},2+n_{x_d,t}, -1)\right],
\label{etat}
\end{align}
where $F$ is the hypergeometric function. We expect this to hold up to rounding errors from the integers. In the following, we compare the exact solution of weights $\braket{\bm w_t}$ from numerically solving Eq.~(\ref{ensemble_dynamics_matrix}) and the binomial solution of weights $\braket{\bm w_t}^{(\rm Bi)}$ in Eq.~(\ref{D_binomial}). 

\begin{figure}
    \centering
    \includegraphics[width = 0.45\textwidth]{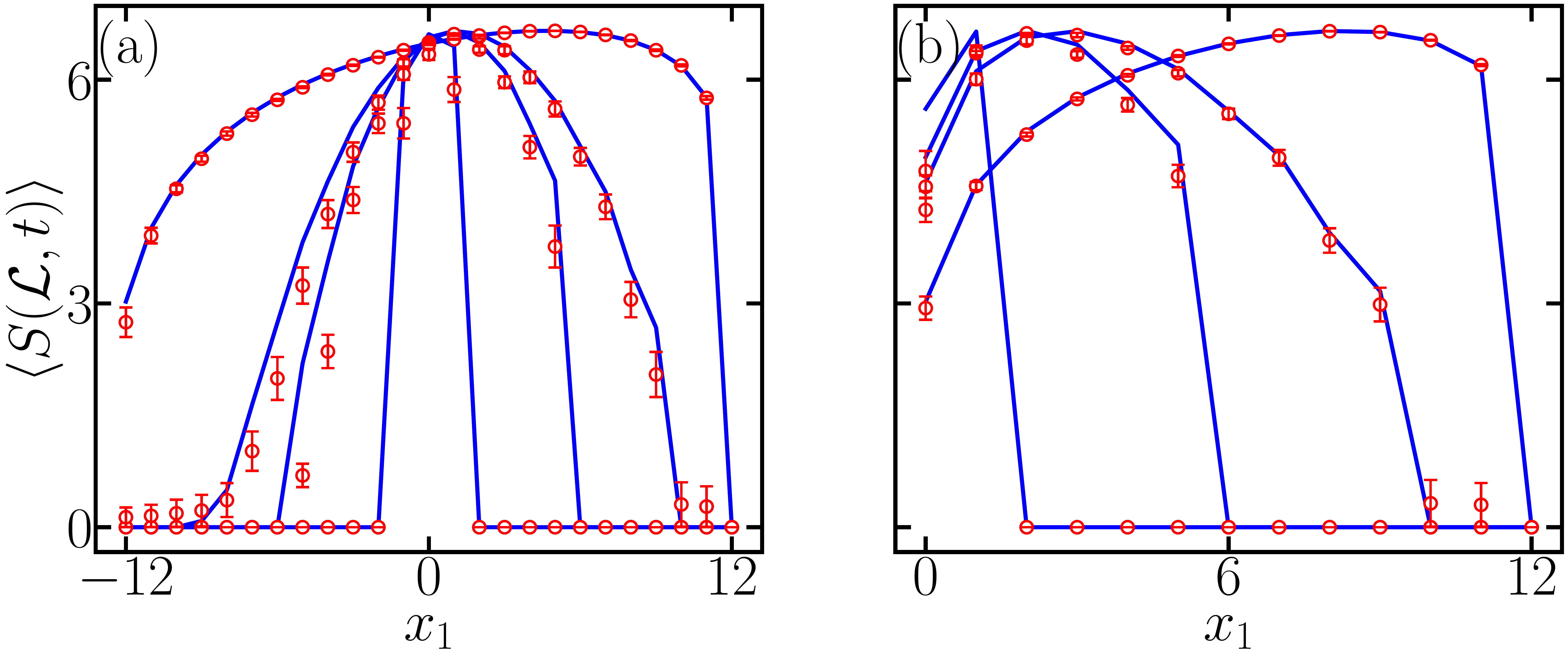}
    \caption{
    Comparison of entropy by real random circuits evolution (red dots with error bars) and the theory from random walk (blue). A squeezer $r=5$ is placed at the center of a 2-D grid system of $25\times 25$ modes. Snapshots are taken at $t=4, 12, 20, \infty$, shown by curves from inside to outside in (a) and left to right in (b). (a) Subsystem $\calL$ is considered to be the region $\calL = [-12, x_1]\times[-12, x_2]$ with $x_2 = x_1$. (b) Subsystem $\calL$ is considered to be the region $\calL = [-|x_1|, |x_1|]\times[-|x_2|, |x_2|]$ with $x_2 = x_1$.
    \label{fig:2D_cross_section}
    }
\end{figure}

As shown in Fig.~\ref{fig:1D_solution_compare}, we see a very good agreement in the 1-D case, up to rounding errors before boundary effects comes in. When there is a finite boundary, standard techniques like image source methods can give more precise solutions. In fact, the continuum limit of Gaussian solutions, as we will present in Sec.~\ref{sec:continuum}, also agrees well with the above results. The perfect agreement of the weights directly indicates the validity of the solution for the entanglement entropy, as demonstrated in Fig.~\ref{fig:1D_entropy_compare}. Note that due to the symmetry among different dimensions, it suffices to consider the 1-D case; however, results in higher dimension reveals more interesting dynamics.

\begin{figure}
    \centering
    \includegraphics[width = 0.475\textwidth]{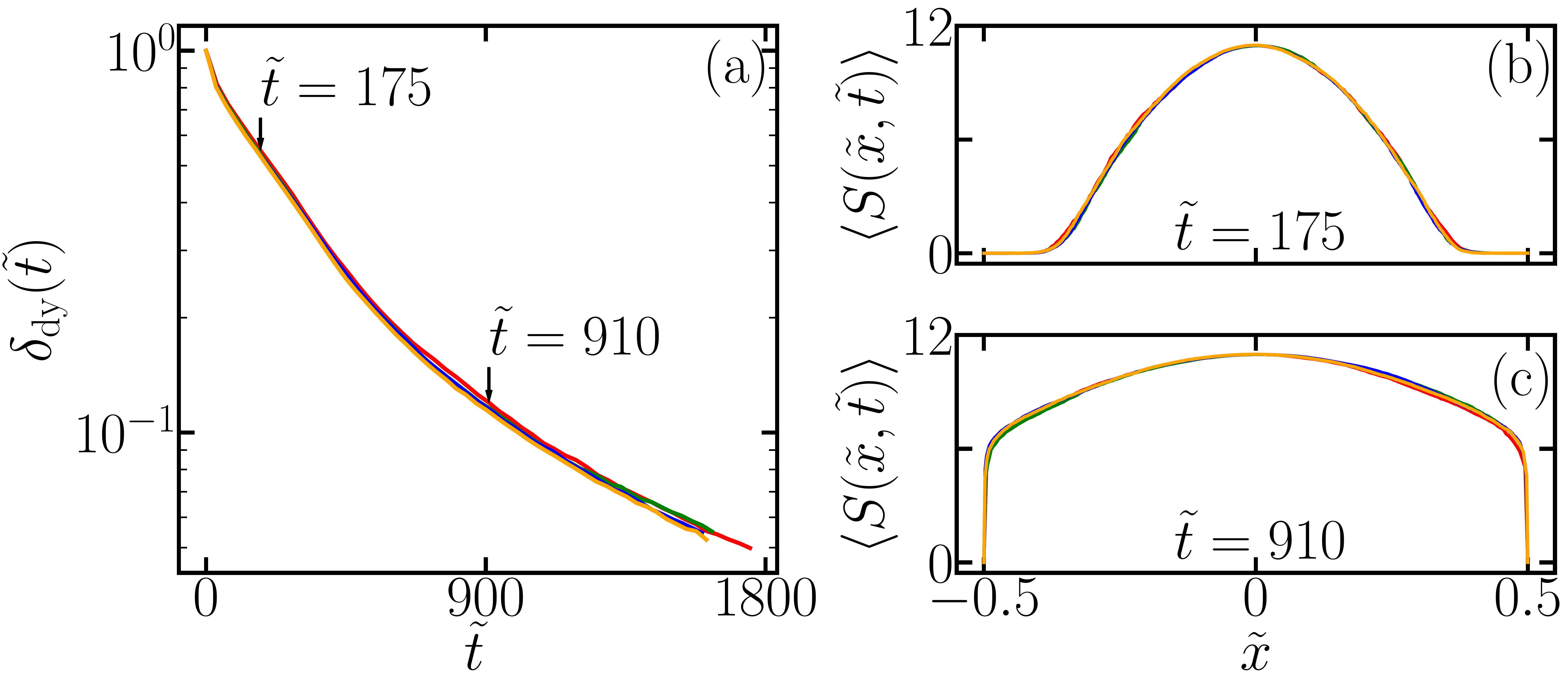}
    \caption{Evolution of the entanglement entropy towards the equilibrium Page curve. A single squeezer with $r=8$ is placed at the center of the system with $M=201, 301, 401, 501$ modes (red, green, blue and yellow). (a) Logarithmic-scale relative deviation $\delta_{\rm dy}(\tilde{t})$ in multiple systems. We re-scale time $\tilde{t} \sim 1/M^2$. (b) and (c) The  entanglement entropy curves of $M=201, 301, 401, 501$ at an effective time corresponding to $t_1=175$ and $t_1=910$ in the system $M=201$.
    \label{fig:dynamic_scale_invariance}
    }
\end{figure}

\begin{figure*}
    \centering
    \subfigure{
    \includegraphics[width = 0.48\textwidth]{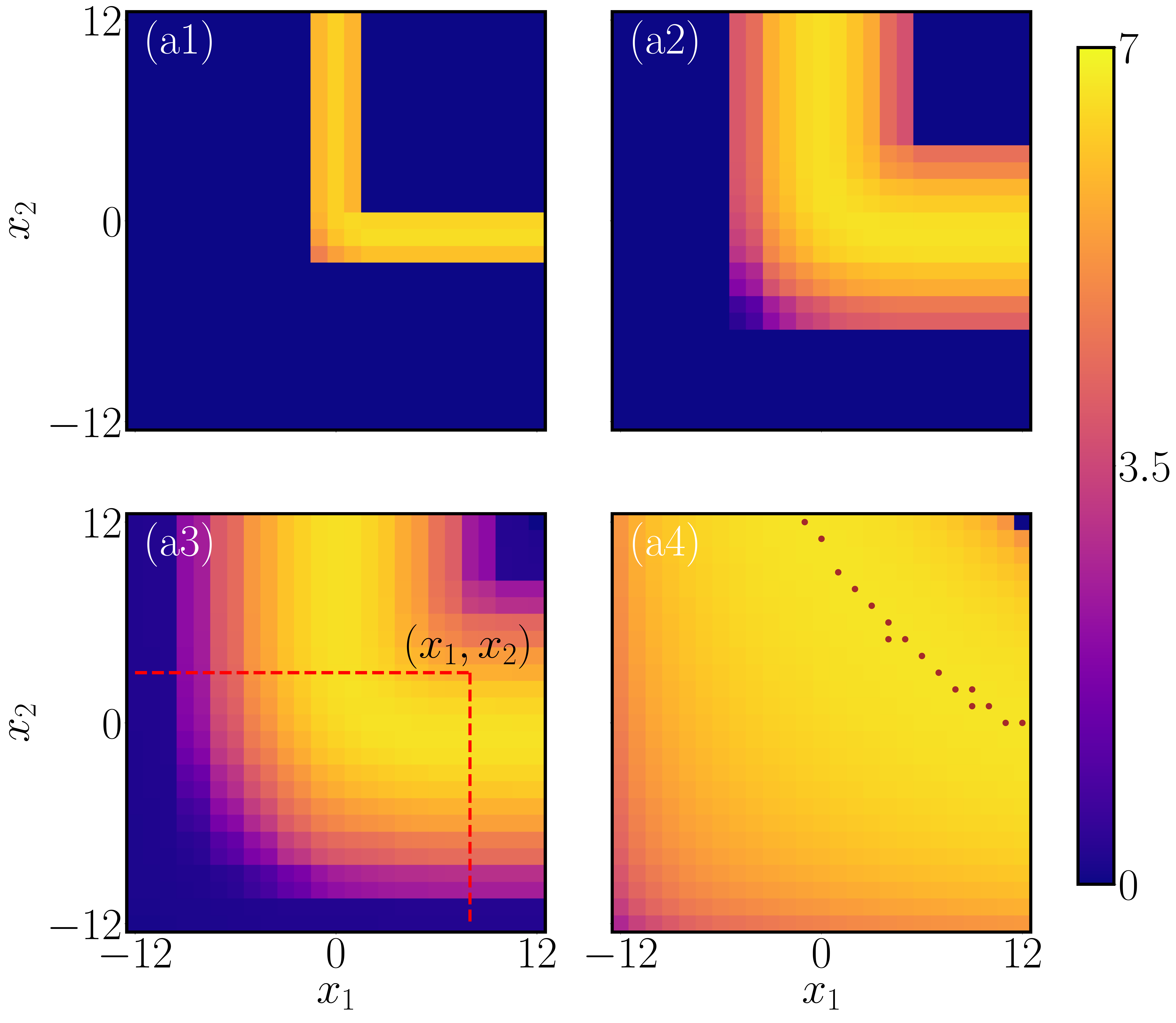}
    }
    \subfigure{
    \includegraphics[width = 0.48\textwidth]{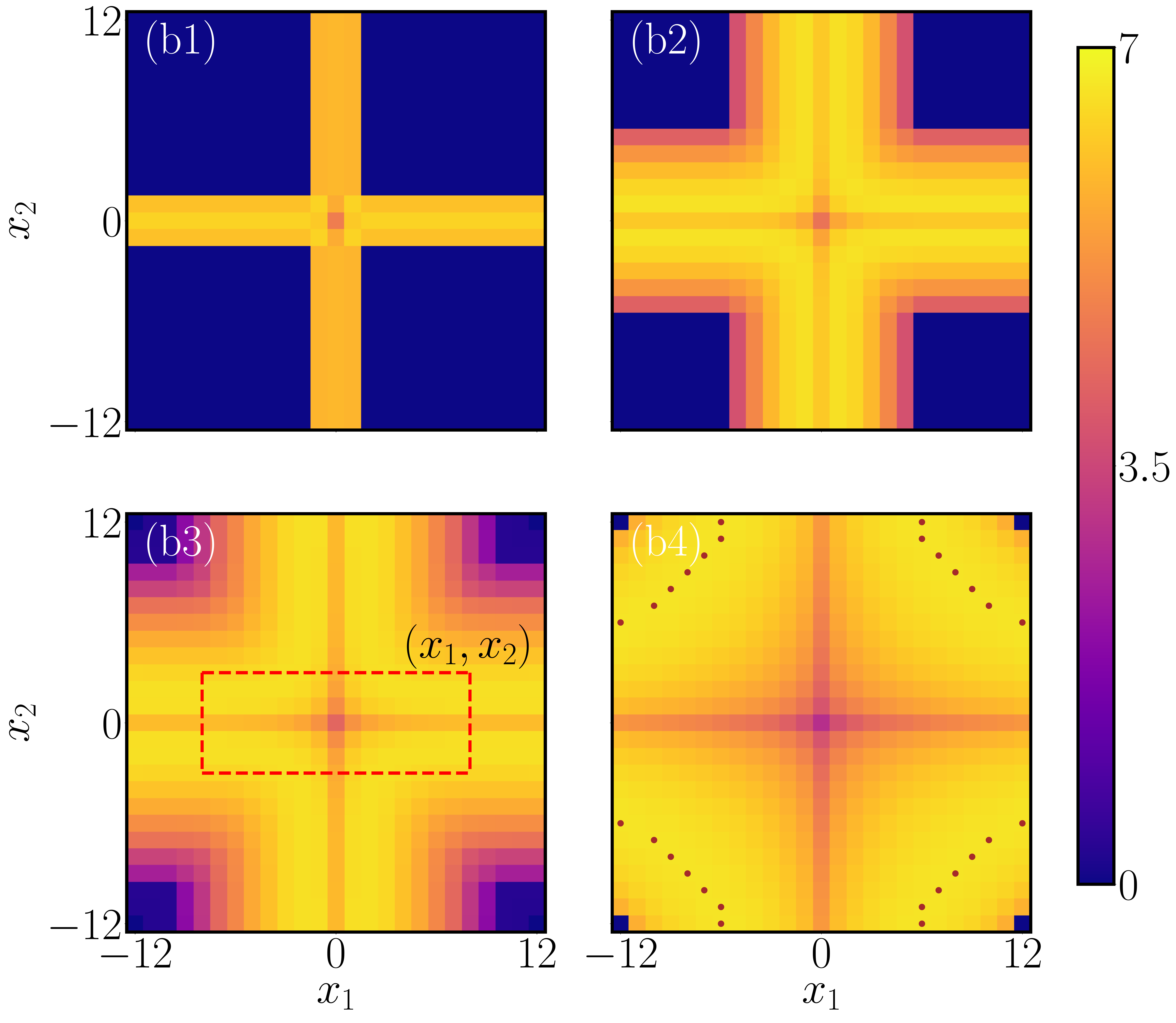}
    }
    \caption{Ensemble-averaged entanglement entropy in a 2-D grid system of $21\times 21$ modes, with a squeezer $r=5$ placed at the origin. The subsystem $\calL=[-12,x_1]\times [-12,x_2]$ is formed from left bottom corner in (a1)-(a4), while the subsystem $\calL = [-|x_1|, |x_1|]\times[-|x_2|, |x_2|]$ is formed from center of system in (b1)-(b4). Red dashed box in (a3) and (b3) shows the two ways to form subsystem $\calL$. Snapshots at $t=4, 12, 20, \infty$ are presented from (1) to (4) in both cases. Brown dots in (a4) and (b4) show the spots with the maximal entanglement in equilibrium state (up to numerical precision). Indeed, in (a4) the equi-entanglement line has the shape of $(x_1+N)(x_2+N)=M^2/2$ while in (b4) it is $4x_1x_2=M^2/2$.
    \label{fig:2D_squares}
    }
\end{figure*}

The 2-D random circuit results are already shown in Fig.~\ref{fig:schematic_2D}, where the region is chosen as each single mode at $(x_1,x_2)$. There, we can see a clear light cone similar to the 1-D case. Here we consider two alternative choices. First, as an analog to the 1-D choice, we can choose the region on the corner $\calL=[-N,x_1]\times [-N,x_2]$, as shown in Fig.~\ref{fig:2D_squares}(a1)-(a4). We see a gradual saturation to the equilibrium, where entanglement entropy along $(x_1+N)(x_2+N)={\rm constant}$ are about equal, as the total weights $\eta_{\calL,t}$ are equal along this line. Second, we can choose $\calL=[-|x_1|,|x_1|]\times [-|x_2|,|x_2|]$ as squares centered at the origin. In Figs.~\ref{fig:2D_squares}(b1)-(b4) We can also see gradual saturation to equilibrium, where the entanglement entropy along $|x_1||x_2|={\rm constant}$ are equal, as the total weights $\eta_{\calL,t}$ are equal along this line. 
To enable comparison in 2-D, we consider cross sections with $x_1=x_2$. As shown in Fig.~\ref{fig:2D_cross_section}, in both choices of the region $\calL$, good agreement between the circuit results and the random-walk results can be seen.

In the above, we see the entanglement entropy from exact ensemble-averaged evolution of Eq.~(\ref{ensemble_dynamics_matrix}) (combined with Eq.~(\ref{SL})) and the actual results from numerical solving the entropy agree well, therefore verifying the underlying random-walk model.

Below, we further address the continuum limit and the entanglement light cone.

\subsubsection{Continuum limit} 
\label{sec:continuum}
We can take the continuum limit of Eq.~(\ref{D_binomial}) and Eq.~(\ref{etat}), which give a $D$ dimensional Gaussian function 
\be
\braket{\bm w_t}^{(\rm Ga)} = \frac{1}{\left(2\pi \left(t/D\right)\right)^{D/2}}\exp\left[-\frac{\|\bm x\|^2}{2(t/D)}\right].
\label{D_gaussian}
\ee
and the corresponds Gaussian error function
\be
\braket{\eta_{\calL,t}}^{(\rm Ga)}
=
\frac{1}{2^D}\prod_{d=1}^D
\left[1+{\rm Erf}\left(\frac{x_d}{\sqrt{2(t/D)}}\right)\right]
\label{etaxt_Gaussian}
\ee 
In Fig.~\ref{fig:1D_solution_compare}, we see a good agreement of the above Gaussian approximation with the other solutions.

The above solutions can be written as a function of $\tilde{\bm x}=\bm x/M$ and $\tilde{t}=t/M^2$ up to normalization, which is well-defined in the continuum limit of $M\to\infty$. 
We verify the continuum limit in 1-D by calculating the deviation measured by the relative 1-norm between the entanglement entropy $\braket{S(\tilde{x},t)}$ and the static value $\braket{S(\tilde{x},\infty)}$ as
\begin{equation}
    \delta_{\rm dy}(t) = \frac{\|\braket{S(\tilde{x},t)}-\braket{S(\tilde{x},\infty)}\|_1}{\|\braket{S(\tilde{x},\infty)}\|_1} \label{eq:delta_rn}
\end{equation}
in the dynamic (short as `dy') process (see Fig.~\ref{fig:dynamic_scale_invariance} (a)), where $\|f(\tilde{x})\|_1=\sum_{\tilde{x}} |f(\tilde{x})|$ sums over the spatial coordinates. We see the re-scaled curves overlap well for systems with different sizes. We can also directly verify in Fig.~\ref{fig:dynamic_scale_invariance} (b)(c) that the entanglement entropy agrees well after re-scaling. It is also worthy to point out that the continuum limit identified for the above single-squeezer case also holds for the multiple-squeezer cases (see Appendix~\ref{app:conti_multi}).

The continuum limit in Eq.~(\ref{D_gaussian}) naturally brings a PDE that descrbes the dynamics
\be 
\partial_t w_{\bm x,t}=\frac{1}{2D}\nabla^2 w_{\bm x,t}.
\label{PDE_weights}
\ee 
The above holds for Cartesian graphs $\calG_D$, in general one needs to adopt the `$\nabla^2$' operator to a graph. Nevertheless, we can obtain observations from $\calG_D$. First, for any region $\calL$, as the continuum limit of Eq.~(\ref{etat_def}), we have $\eta_{\calL,t}= \int_\calL d\bm x w_{\bm x,t} $, therefore the time derivative
\be 
\partial_{t} \eta_{\calL,t}=\frac{1}{2D}\int_\calL d\bm x \nabla^2 w_{\bm x,t}=\frac{1}{2D}\int_{\partial \calL}  d\ell \hat{\bm n}_{\bm x}\cdot \nabla w_{\bm x,t}
\label{PDE_eta}
\ee 
becomes a loop integral on the boundary $\partial \calL$ of the `flow' $\nabla w_{\bm x,t}$ along the normal direction $\hat{\bm n}_{\bm x}$. As the continuum limit of Eq.~(\ref{eta_boundary}), it shows that the entanglement entropy's dynamics is governed by the boundary.

The above PDEs~(\ref{PDE_weights}) and~(\ref{PDE_eta}) describe the evolution of weights, one relies on Eq.~(\ref{SL}) to connect to the entanglement evolution. Alternatively, one can directly focus on the entanglement entropy and design a coupled nonlinear diffusive epidemiology model to describe the entanglement dynamics, as we present in Appendix~\ref{app:epidemiology}. We note that both nonlinear diffusion equations~\cite{nahum2017quantum} and epidemiology models~\cite{qi2018quantum} have been separately used in modeling quantum information scrambling. This phenomenological model shows an interesting combination of both to describe a unique CV entanglement growth process.

\subsubsection{Entanglement light cone and sudden growth}
\label{sec:light_cone}
We now proceed to identify unique phenomena for the entanglement evolution. One of such an evolution is depicted in Fig.~\ref{fig:lightcone}(a), where we see entanglement diffusively spreads from the source at the origin. We can introduce an entanglement light cone (green lines) and a wave front of the entanglement sudden growth (black lines), as will be explained in the following paragraphs.

\begin{figure}
    \centering
    \includegraphics[width = 0.475\textwidth]{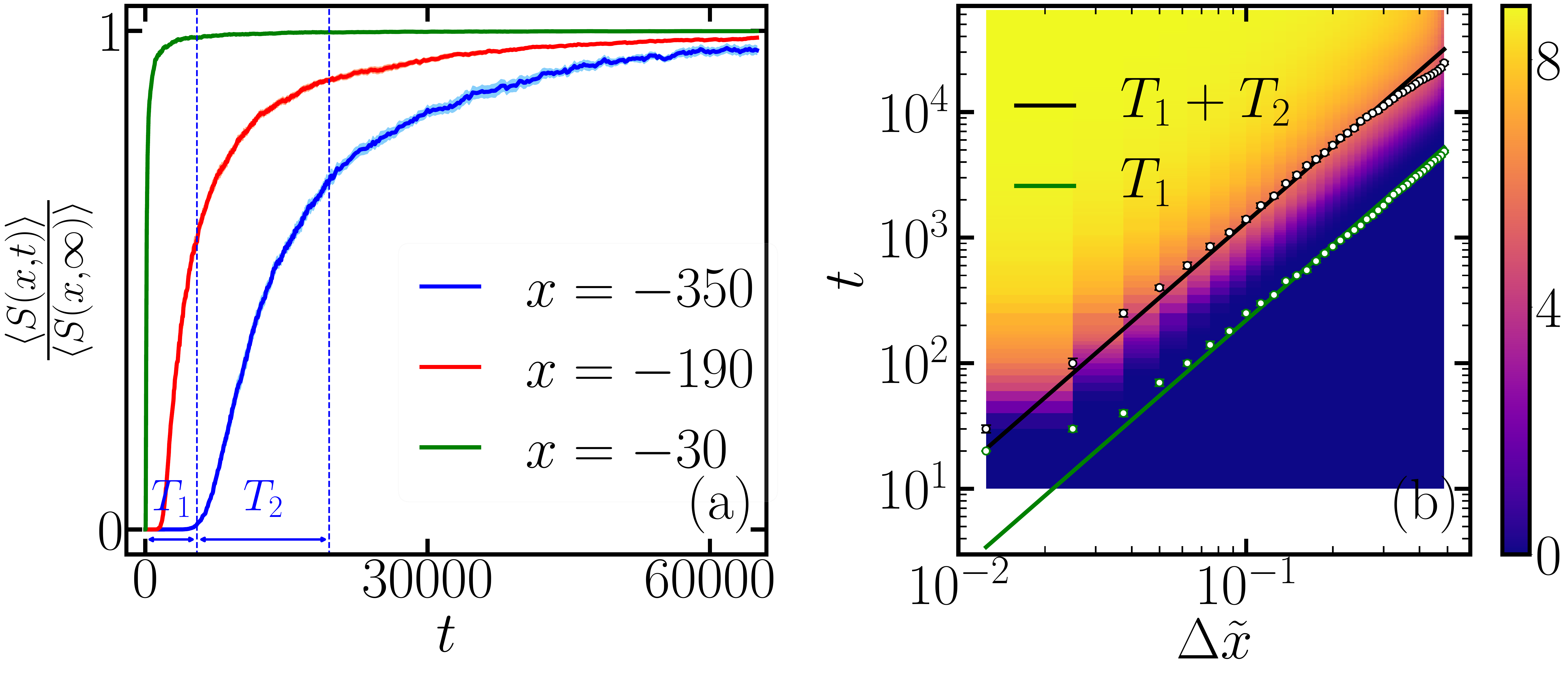}
    \caption{Entropy time evolution of spots in a system of $M=801$ modes. Squeezer $r=6$ is placed at the center of system. Thresholds for determining $T_1$ and $T_2$ is 1\% and 70\% of the static value. 
    (a) Time evolution of spots. $T_1$ and $T_2$ for the mode at $x=-350$ are labeled. Shadow areas show the standard deviation of the time evolution curves.
    (b) Dependence of $T_1+T_2$ and $T_1$ on their relative distance to the squeezer $\Delta \tilde{x}$ represented by black and green lines. Both $t$ and $\Delta\tilde{x}$ are plotted in a logarithmic scale, with sub-ticks indicating numbers with an equal internal in a linear scale. Solid lines represent the fitting results $\sim (\Delta \tilde{x})^2$. The dots represent $T_1,T_2$ for form the ensemble averaged entropy $\braket{S(x,t)}$, while the error bar indicates the precision due to finite sample size.
    The background heatmap shows the entropy distribution $\braket{S(x,t)}$ among the system.
    \label{fig:spot_evo_M801}
    }
\end{figure}

To further understand the dynamics, we focus on particular modes and consider $\braket{S(x,t)}$ as a function of $t$ (see Fig.~\ref{fig:spot_evo_M801}(a) for examples). We observe a three-stage evolution: (1) In the first period $0\le t < T_1$, the entanglement $\braket{S(x,t)}$ is almost zero. This is the time period before the entanglement light cone reaches location $x$. (2) In the second period $T_1 \le t < T_1+T_2$, the entanglement light cone reaches the spot and causes a rapid increase in $\braket{S(x,t)}$, after which $\braket{S(x,t)}$ gets close to $\braket{S(x,\infty)}$. (3) In the last period $T_1+T_2 \le t $, $\braket{S(x,t)}$ gradually saturates towards $\braket{S(x,\infty)}$. We numerically investigate the scaling of $T_1, T_2$ with respect to the distance $\Delta \tilde{x}$ to the initial squeezer, which reveals the physics of entanglement growth.

In the first period $t<T_{1}$, the entanglement entropy at each spot is negligible. As we see in Fig.~\ref{fig:lightcone}(a), the green curves show the threshold $T_1$ for spots at different distances $\Delta \tilde{x}$---a parabolic entanglement light cone much slower than the usual linear light cone of operator spreading. The second period $T_1\le t <T_1+T_2$ describes the wave-front of entanglement rapid growth. As we see in Fig.~\ref{fig:lightcone}(a), the black curve depicts the threshold $T_1+T_2$ for spots at different distances $\Delta \tilde{x}$. The parabolic shape again indicates a diffusion behavior. 

This parabolic light cone can be explained by our statistical theory. We want a constant fraction of the maximum entanglement $\epsilon S_0(r)=S(\eta_{\calL,t})$ in Eq.~(\ref{SL}), combining with Eq.~(\ref{etat}) or Eq.~(\ref{etaxt_Gaussian}) we can solve $T_1, T_1+T_2$ precisely, despite the analytical formula being lengthy, one immediately recognizes the scaling 
\be 
T_1, T_2 \sim (\Delta \tilde{x})^2 f(r),
\label{eq:T1T2}
\ee 
with some function $f$ of the squeezing strength $r$.
Indeed, as shown by Fig.~\ref{fig:spot_evo_M801}(b), the green curve (entanglement light cone) and the black curve (entanglement sudden growth) both agrees well with the quadratic fitting. This is indeed consistent with the OTOC diffusion identified in~\cite{zhuang2019scrambling}, revealing an unique universal behavior intrinsic to CV quantum networks and absent in DV circuits~\cite{nahum2017quantum,nahum2018operator}.

\subsection{Growth of multi-partite entanglement measured by distributed sensing}
\label{sec:multipartite_E}

\begin{figure}
    \centering    
    \includegraphics[width =0.4\textwidth]{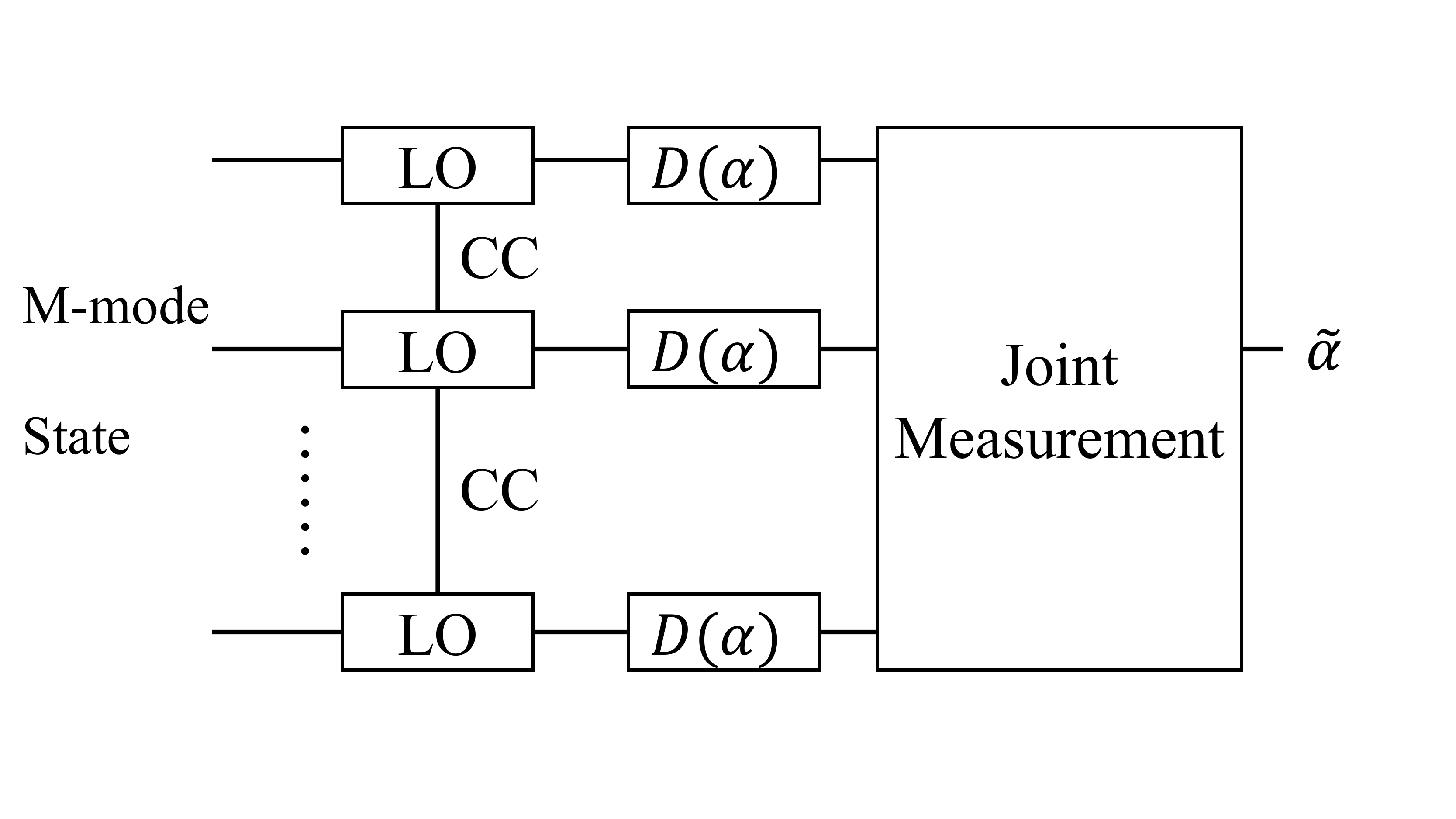}
    \caption{Schematic of the sensing process for multipartite entanglement witness. LO: local operations. CC: classical communications. After the energy-preserving LOCC, each mode of the input state goes through a displacement $D(\alpha)$; An estimator of the displacement amplitude $\tilde{\alpha}$ is generated from a joint measurement. 
    \label{fig:witness}
    }
\end{figure}

So far we have focused on bipartite entanglement between a subsystem $\calL$ and its complement $\calR$. In quantum networks, many applications often require multipartite entanglement, which is in general difficult to characterize~\cite{adesso2014}. Here we take an operational approach from a quantum sensing perspective.
An important application of the entanglement generated in such a random quantum network is distributed sensing~\cite{zhuang2018distributed,zhuang2019physical,guo2020distributed,xia2019entangled}, where multi-partite entanglement enables an improvement in the measurement sensitivity. In the case of measuring uniform real displacements of amplitude $\alpha$ on all modes, one can prove that considering a total mean photon number $|\calG|N_S$, the optimum $|\calG|$-mode separable state can only offer a variance 
\be 
V_C=\frac{1}{4}\frac{1}{|\calG|(\sqrt{N_S+1}+\sqrt{N_S})^2}\sim \frac{1}{16|\calG|N_S}
\ee 
in estimating the displacement $\alpha$ (the standard quantum limit). Therefore beating the above precision limit is an evidence of entanglement. In fact, one can show that the optimum precision attainable by {\it all} entangled state is
\be 
V_E=\frac{1}{4}\frac{1}{|\calG|(\sqrt{|\calG|N_S+1}+\sqrt{MN_S})^2}\sim  \frac{1}{16|\calG|^2N_S},
\ee 
which possesses the Heisenberg scaling of $V_E\sim 1/|\calG|^2$ that is only possible with multipartite entanglement. Therefore, we can define an entanglement witness for any $|\calG|$-mode state $\rho$ with total energy $|\calG|N_S$ as 
\be
\calE(\rho)=\max_{{\rm LOCC}^\prime}\log_2\left(V_C/V(\rho)\right),
\ee 
where $V(\rho)$ is variance achievable by performing a local-operations (LO) and classical communications (CC) on the input $\rho$ (hence ${\rm LOCC}^\prime$), while keeping the total energy conserved (see Fig.~\ref{fig:witness}). We maximize over all such LOCC schemes. We have $\calE(\rho)= 0$ for all separable states~\cite{Note4}, and $\calE(\rho)\le \log_2(V_C/V_E)\sim \log_2|\calG|$ for all states.

For the state $\rho(t)$ generated in the single-squeezer random network at time step $t$, we can design an estimator to obtain a lower bound of $\calE(\rho(t))$. Given the weight $\{w_{\bm x,t}\}$ on distributing the SV state to $|\calG|$ modes, we can design the following measurement protocol.
First, we perform a phase rotation on each mode such that the displacements act on the corresponding squeezed quadrature; then we perform homodyne measurements on the corresponding quadratures to obtain the results $\{\tilde{\alpha}_{\bm x,t}\}_{\bm x\in \calG}$. The estimator $\tilde{\alpha}=\sum_{\bm x\in \calG} \sqrt{w_{\bm x,t}} \tilde{\alpha}_{\bm x,t}/\sum_{\bm x\in \calG} \sqrt{w_{\bm x,t}}$, which gives the variance
\be 
V(t)=\frac{1}{4}\frac{1}{\overline{|\calG|}(\sqrt{|\calG|N_S+1}+\sqrt{|\calG|N_S})^2},
\ee 
where the effective number of modes $\overline{|\calG|}=(\sum_{\bm x \in \calG} \sqrt{w_{\bm x,t}})^2\in[1,|\calG|]$ that are entangled provides the advantage.

Combing the weights in Eqs.~(\ref{D_binomial}) and (\ref{D_gaussian}), we can obtain the effective entangled mode number for the $D$ dimensional Cartesian graph $\calG_D$
\be
\overline{|\calG|}\simeq \int d^{D}\bm x\braket{\bm w_t}^{(\rm Ga)}
= \left(2(2\pi t/D)^{1/2}\right)^D
\ee 
which leads to the entanglement witness
\be
\calE(\rho(t))\ge  \log_2\left(V_C/V(t)\right)\simeq \frac{D}{2}\log_2(8\pi t/D),
\ee 
before the boundary effect comes in, when the effective modes become comparable to $|\calG|$ at $t\sim |\calG|^2$.

\section{Multiple squeezers: sparse limit}
\label{sec:multiple_squeezers}

\begin{figure}
    \centering
    \includegraphics[width = 0.45\textwidth]{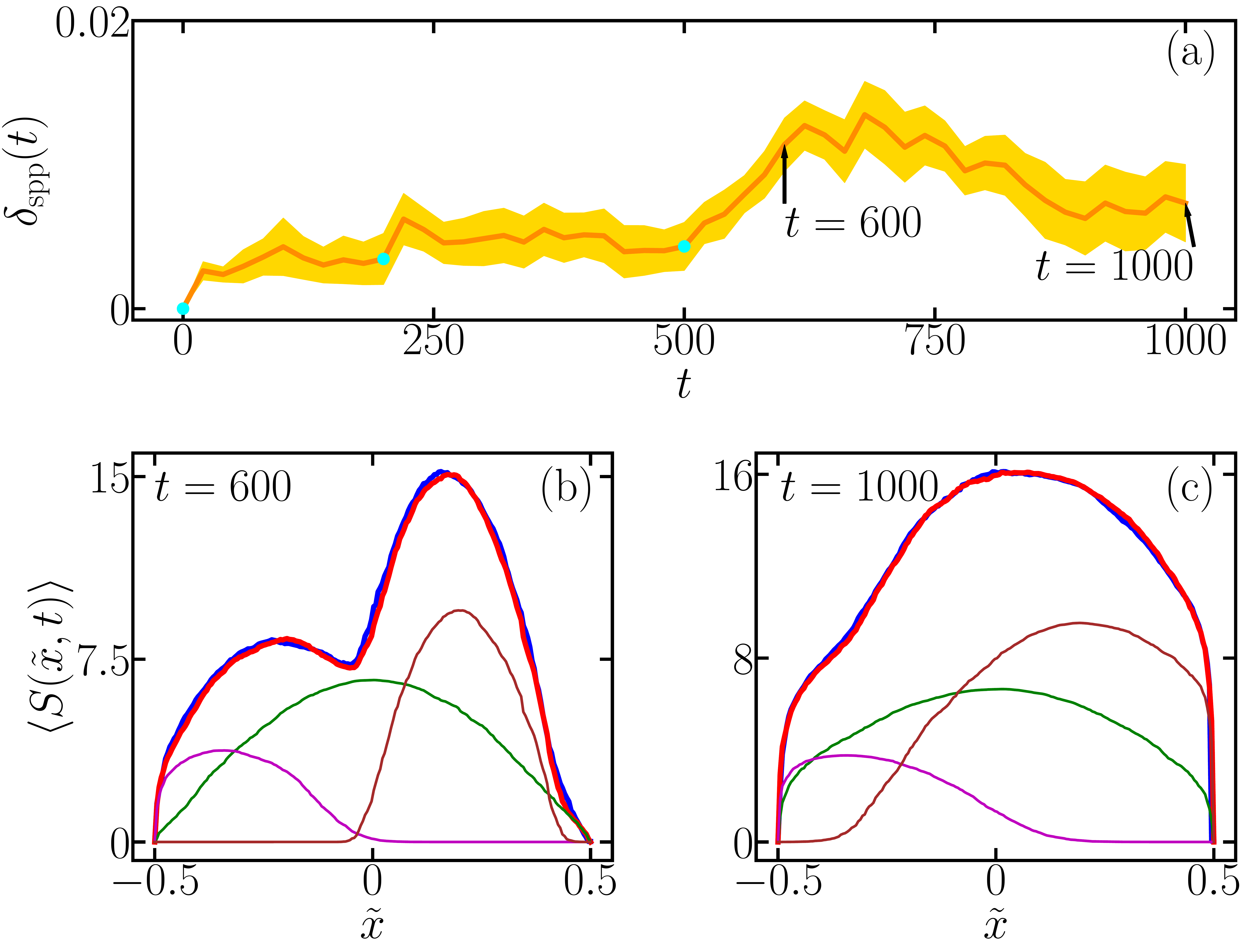}
    \caption{Superposition in the three-squeezer case of Fig.~\ref{fig:lightcone}(b). (a) Relative deviation of linear superposition. The orange line is the mean and the yellow area shows the precision due to finite sample size. Cyan dots show the time when the three squeezers are applied accordingly. Arrows represent the time when snapshots in (b) and (c) are taken. (b), (c) Snapshots of entanglement entropy curves at $t=600, 1000$. Blue and red lines are  $\braket{S(\tilde{x},t)}$ and superposition $\braket{S_{\rm spp}(\tilde{x},t)}$; while green, purple and brown ones show $\braket{S_i(\tilde{x},t)}$ generated by each of the three squeezers. 
    \label{fig:dynamic_spacetime_spp_N3}
    }
\end{figure}

In Sec.~\ref{sec:statistical_theory}, we focus on random networks with a single squeezer and present a thorough theory for the entanglement dynamics and equilibrium, via an exact mapping to random walk on a graph. Quantum networks are likely to have multiple squeezers; therefore, we extend our analyses to random networks with multiple squeezers in this section. A surprising linear superposition law is numerically observed and theoretically explained.

We begin with an intuitive example of three squeezers in 1-D Cartesian graph in Fig.~\ref{fig:lightcone}(b). The overall evolution of the entanglement entropy looks like a linear superposition of three independent squeezers, despite the nonlinearity of the entanglement dynamics. Following this observation, we consider a random circuit $\mathbb{C}$ with $N_q$ squeezers at different space-time coordinates $\{\bm \xi_k^\star=(\bm x_k^\star,t_k^\star)\}_{k=1}^{N_q}$, with squeezing strengths $\{r_k\}_{k=1}^{N_q}$. Linear superposition (spp) means the entanglement entropy of subsystem $\calL$, $\braket{S(\calL,t)}\simeq  \braket{S_{\rm spp}(\calL,t)}$, where
\begin{equation}
   \braket{S_{\rm spp}(\calL,t)} = \sum_{k=1}^{N_q} \braket{S_k(\calL,t)}
   \label{eq:superposition}
\end{equation}
is a simple sum of the ensemble averages $\braket{S_k(\calL,t)}$. Here $S_k(\calL,t)$ is generated from a random circuit $\mathbb{C}_k$ with a single squeezer of strength $r_k$ at $\bm \xi_k^\star$, therefore can be calculated by the random-walk mapping in Eqs.~(\ref{ensemble_dynamics_matrix}) and~(\ref{SL}). Note that the random beamsplitters in all $N_q$ single-squeezer circuits $\{\mathbb{C}_i\}_{k=1}^{N_q}$ and the original circuit $\mathbb{C}$ are independent. To test linear superposition, we numerically calculate the deviation per mode
\be 
\Delta S_{\rm spp}(t) = \frac{1}{|\calG|}\|\braket{S(\calL,t)}-\braket{S_{\rm spp}(\calL,t)}\|_1.
\label{eq:ab_deviation_spp}
\ee 
To evaluate the relative deviation, we can also rescale the deviation relative to the steady state value, as $\delta_{\rm spp}(t)=\Delta S_{\rm spp}(t)/(\|\braket{S(\calL,\infty)}\|_1/|\calG|)$. Both deviations are system-size independent in the continuum limit.

In Fig.~\ref{fig:dynamic_spacetime_spp_N3}(a), we evaluate the deviation for the 1-D three-squeezer case considered in Fig.~\ref{fig:lightcone}(b). We see that the relative deviation $\delta_{\rm spp}(t)$ is small ($<2\%$) through the entire dynamical evolution. To be more explicit, in Figs.~\ref{fig:dynamic_spacetime_spp_N3}(b) and (c), we directly plot $\braket{S(\tilde{x},t)}$ (blue) at various times, which agrees well with the superposition result $\braket{S_{\rm spp}(\tilde{x},t)}$ (red). 


Following the above observation, Sec.~\ref{sec:spp_theory} provides a theory at the sparse squeezers limit, which predicts linear superposition for both the equilibrium Page curves and the dynamical evolution; These two aspects are then investigated in Sec.~\ref{sec:Page} and Sec.~\ref{sec:superposition}.

\subsection{Theory of the sparse squeezers limit}
\label{sec:spp_theory}
Inspired by the above numerical findings in 1-D, we present the following theory to explain the linear superposition law. For the circuit $\mathbb{C}$, similar to Eq.~(\ref{axt}), each mode at $\bm x\in \calG$ and time $t$ can be written as
\be 
a_{\bm \xi}=\sum_{k=1}^{N_q}e^{i\theta_{\bm \xi;k}}\sqrt{w_{\bm \xi;\bm \xi_k^\star}} a_{\rm SV;k}+{\rm vac},
\label{axt_multi}
\ee 
where we use the simplified notation $\bm \xi=(\bm x,t)$. Since the squeezers' locations $\bm \xi_k^\star$ are different, all fully random phases $\{\theta_{\bm \xi;k}\}$ are independent. Each set of positive weights $\{w_{\bm \xi;\bm \xi_k^\star}\}_{\bm x\in \calG}$ describes the energy-splitting among the modes of each single-mode SV $a_{\rm SV;k}$. From these weights, it immediately follows that a total portion $\eta_{\calL,t;k}=\sum_{\bm x\in \calL}  w_{\bm \xi;\bm \xi_k}$ of each single mode squeezer end up in subsystem $\calL$. 

\begin{figure}
    \centering
    \includegraphics[width = 0.475\textwidth]{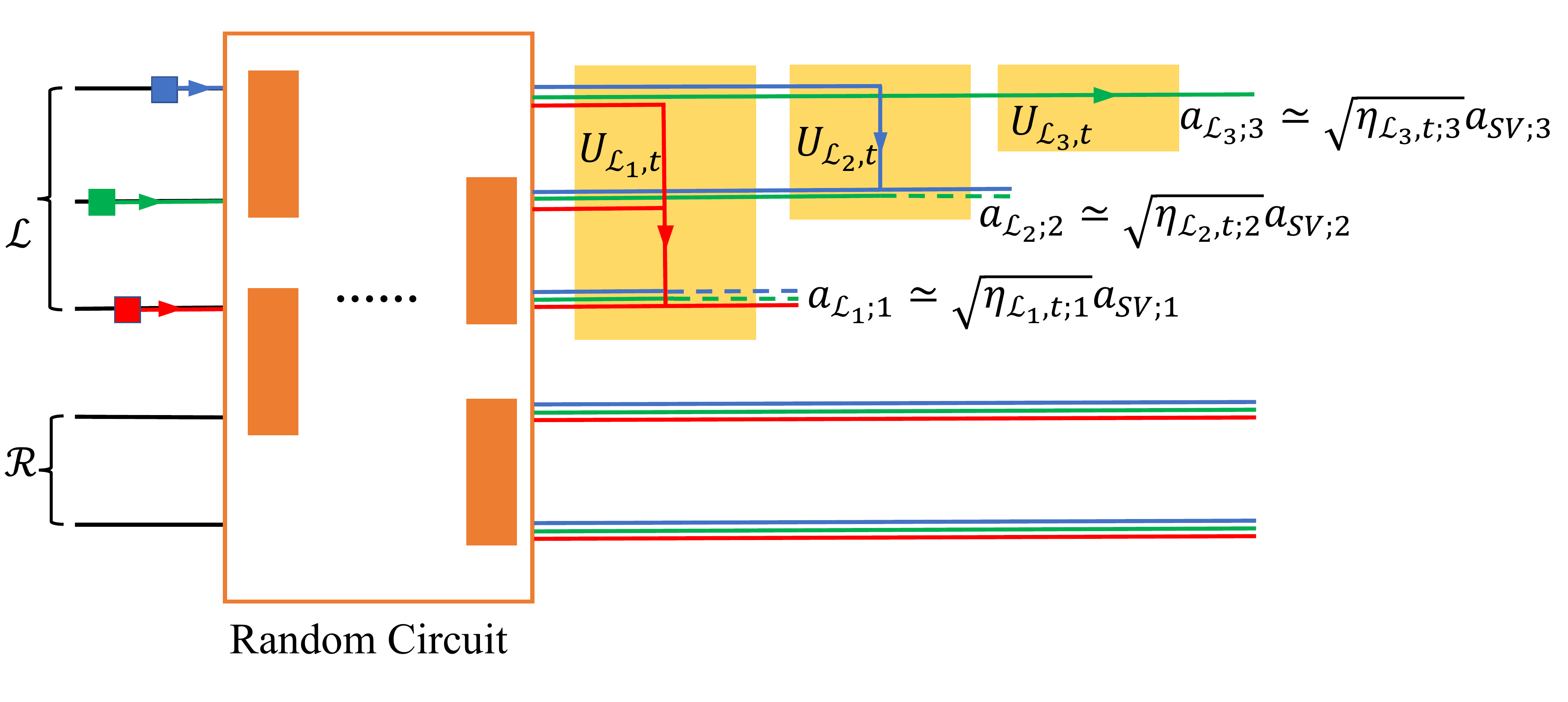}
    \caption{Schematic of the multi-mode transforms. Solid lines with color red, green and blue represent energy flow from three squeezed mode at different space-time. After the random circuit, we apply a set of unitaries $U_{\calL_s,t}$ to concentrate the squeezed modes into each single modes $a_{\calL_s;s}$. The dashed lines represent residues from other squeezed modes after passing through the unitary transform $U_{\calL_s,t}$. The residue of other squeezing modes can be neglected in the sparse-squeezing limit. 
    \label{fig:schematic_transform_multi}
    }
\end{figure}

To evaluate the entropy of $\calL$, similar to Sec~\ref{sec:mapping}, we consider a set of passive linear optics unitaries $\{U_{\calL_{s},t}\}_{s=1}^{N_q}$ to manipulate the power distribution of the SV within $\calL$ (see Fig.~\ref{fig:schematic_transform_multi}). The first transform $U_{\calL_1,t}$ acts on the entire system $\calL_1=\calL$ to concentrate the first SV part to the mode at an arbitrary mode $\bm y_1\in \calL_1$ as
\be 
a_{\calL_1;1}=\sqrt{\eta_{\calL_1,t;1}}a_{\rm SV;1}+\sum_{k=2}^{N_q}e^{i\theta_{\bm \xi;k}^{(1)}}\sqrt{w_{(\bm y_1,t);\bm \xi_k}^{(1)}} a_{\rm SV;k}+{\rm vac},
\ee 
where $\eta_{\calL_1,t;1}=\sum_{\bm x\in \calL_1} w_{\bm \xi; \bm \xi_k}$ is the total portion of $a_{\rm SV;1}$ in the subsystem $\calL_1$. For the first transform, we have $\eta_{\calL_1,t;1}= \eta_{\calL,t;1}$. We can denote the remaining $(M-1)$ modes after the transform as $\calL_2$, with the weights $\{w_{\bm \xi;\bm \xi_k}^{(1)}\}_{\bm x\in \calL}$ for squeezers $\{a_{\rm SV;k}\}_{k= 2}^{N_q}$. Note that a $w_{(\bm y_1,t);\bm \xi_k}^{(1)}$ portion of each squeezer $a_{\rm SV;k}$ ($k\ge2$) is also mixed in the mode $a_{\calL_1;1}$, and not in the new subsystem $\calL_2$.

After the first $(s-1)$ transforms ($N_q\ge s\ge2$), we have subsystem $\calL_{s}$ with $|\calG|-s+1$ modes . The $s$-th transform $U_{\calL_s,t}$ acts on the subsystem $\calL_{s}$ to concentrate the power of the SV $a_{\rm SV; s}$ to the mode at $\bm y_s\in \calL_s$ as
\be 
a_{\calL_s;s}=\sqrt{\eta_{\calL_{s},t;s}}a_{\rm SV; s}+\sum_{k=s+1}^{N_q} 
e^{i\theta_{\bm \xi;k}^{(s)}}\sqrt{w_{(\bm y_s,t);\bm \xi_k}^{(s)}} a_{\rm SV;k}+{\rm vac},
\ee 
where $\eta_{\calL_{s},t;s}=\sum_{\bm x\in \calL_s} w_{\bm \xi; \bm \xi_k}^{(s-1)} \le \eta_{\calL,t;s}$ is total power portion of $a_{\rm SV; s}$ in the subsystem $\calL_s$.

\begin{figure}
    \centering
    \subfigure{
    \includegraphics[width = 0.165\textwidth]{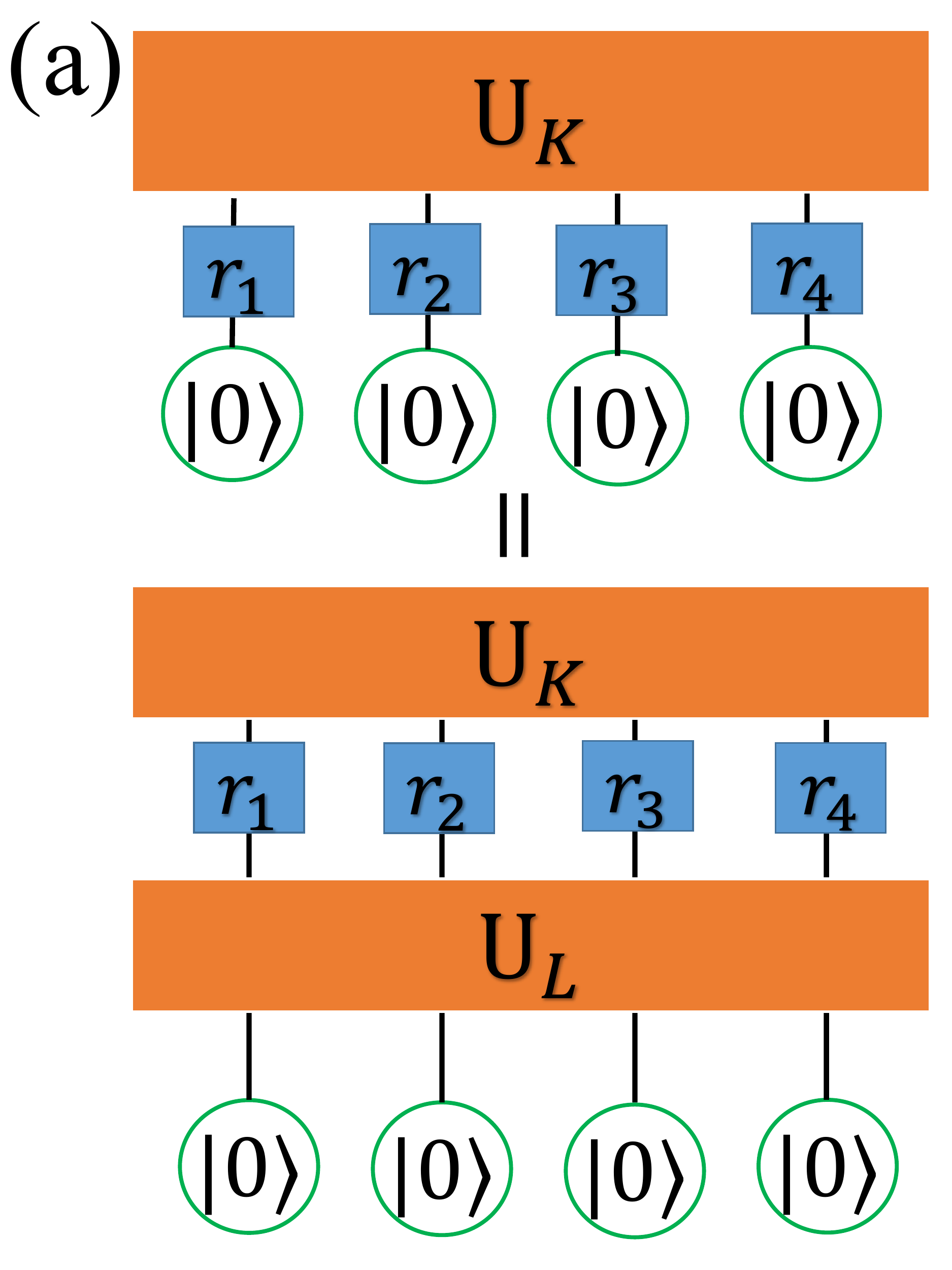}
    \label{fig:schematic_Page}}
    \subfigure{
    \includegraphics[width = 0.26\textwidth]{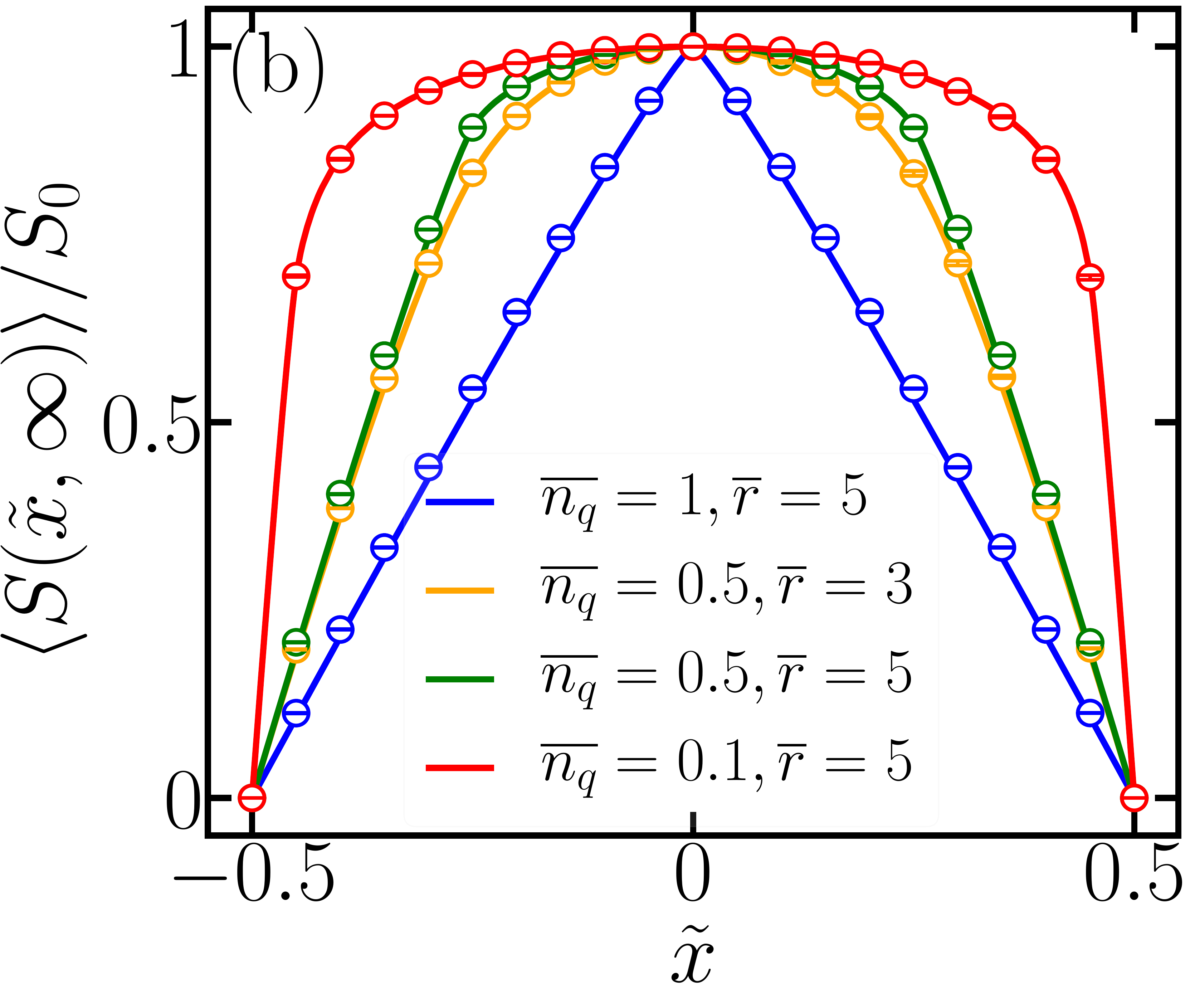} \label{fig:static_scale_invariance}}
    \caption{
    (a) Schematic of the circuit generating the CV Page curves. Because the passive transform $U_{\bm K}$ acting on vacuum still produces vacuum, the Euler decomposition (bottom) on vacuum input is equivalent to applying the passive transform $U_{\bm K}$ after a layer of squeezers with strength $\{r_k\}_{k=1}^{N_q}$. 
    (b) 
    Re-scaled Page curves and their dependence on ${\overline{n_q}, \overline{r}}$ in the system $M = 400$. Solid lines are for identical squeezers with $r_i = \overline{r}$. 
    Scatter circles show the Page curve with randomly chosen $N_q$ squeezers of the same $\overline{r}$ (20 points are shown as an example). Error bars inside circles show fluctuations among the random configurations of the squeezing strengths. Results for different system size $M$ are identical, similar to (b).
    \label{fig:pages}
    }
\end{figure}

After all transforms $\{U_{\calL_{s},t}\}_{s=1}^{N_q}$, we obtain modes $\{a_{\calL_s;s}\}_{s=1}^{N_q}$, which contains all the SVs in $\calL$~\cite{Note1}; while the rest of the modes are all in vacuum. Due to the unitarity of all transforms, the entropy of the original system $S(\calL,t)=S\left(\{a_{\calL_s;s}\}_{s=1}^{N_q}\right)$ equals the entropy of these modes. To further evaluate the entropy, we impose two constraints---independence of weights and $N_q\ll |\calG|$---on the random circuit to enable an approximate solution. 

First, we assume the weights $\{w_{\bm \xi;\bm \xi_k^\star}\}_{\bm x\in \calG}$  are independent between different squeezers (indexed by $k$). This assumption is true in two scenarios:
(1) when any two squeezers are far away from each other in terms of the shortest path on the graph and/or the time difference.
In this case, the independence of weights hold at any time $t$;
(2) as long as the number of squeezers $N_q\ll |\calL|$ and the squeezers do not act on top of each other, the correlations between the weights will be small at late time. This is especially true for the Page curve $\braket{S(\calL,\infty)}$.
The independence of the weights means that after the $s$-th passive linear unitary, the new weights $\{w_{\bm \xi;\bm \xi_k}^{(s)}\}_{\bm x\in \calL}$ for the remaining squeezers are still randomly distributed across the entire system (not concentrated on any mode). 

Second, we assume that the total number of squeezers is small, i.e., $N_q\ll |\calG|$. This allows us to approximate the weights concentrated from subsystem $\calL_{s}$ as the global total weights, i.e.,
$
\eta_{\calL_{s},t;s}\simeq \eta_{\calL,t;s}, 1\le s \le N_q.
$
Therefore, we conclude that after the set of passive unitaries $\{U_{\calL_{s},t}\}_{s=1}^{N_q}$, we can approximately obtain a set of modes 
\be 
\{a_{\calL;s}\equiv  \sqrt{\eta_{\calL,t;s}}a_{\rm SV;s}+{\rm vac}\}_{s=1}^{N_q},
\ee 
which is a product of lossy single-mode SVs. From additivity of entropy, we arrive at linear superposition $\braket{S(\calL,t)}\simeq  \braket{S_{\rm spp}(\calL,t)}$, note that each single-squeezer term can also be written out explicitly as $S(\eta_{\calL,t;s})$ from Eq.~(\ref{SL_preliminary}) by replacing $r$ with $r_k$.

Note that linear superposition law holds not only for the dynamical evolution at any $t$ (as systematically explored in Section~\ref{sec:superposition}), but also for the equilibrium Page curves at $t=\infty$ (see Section~\ref{sec:Page}).

\subsection{Page curves with multiple squeezers}
\label{sec:Page}

\begin{figure}
    \centering
    \includegraphics[width = 0.45\textwidth]{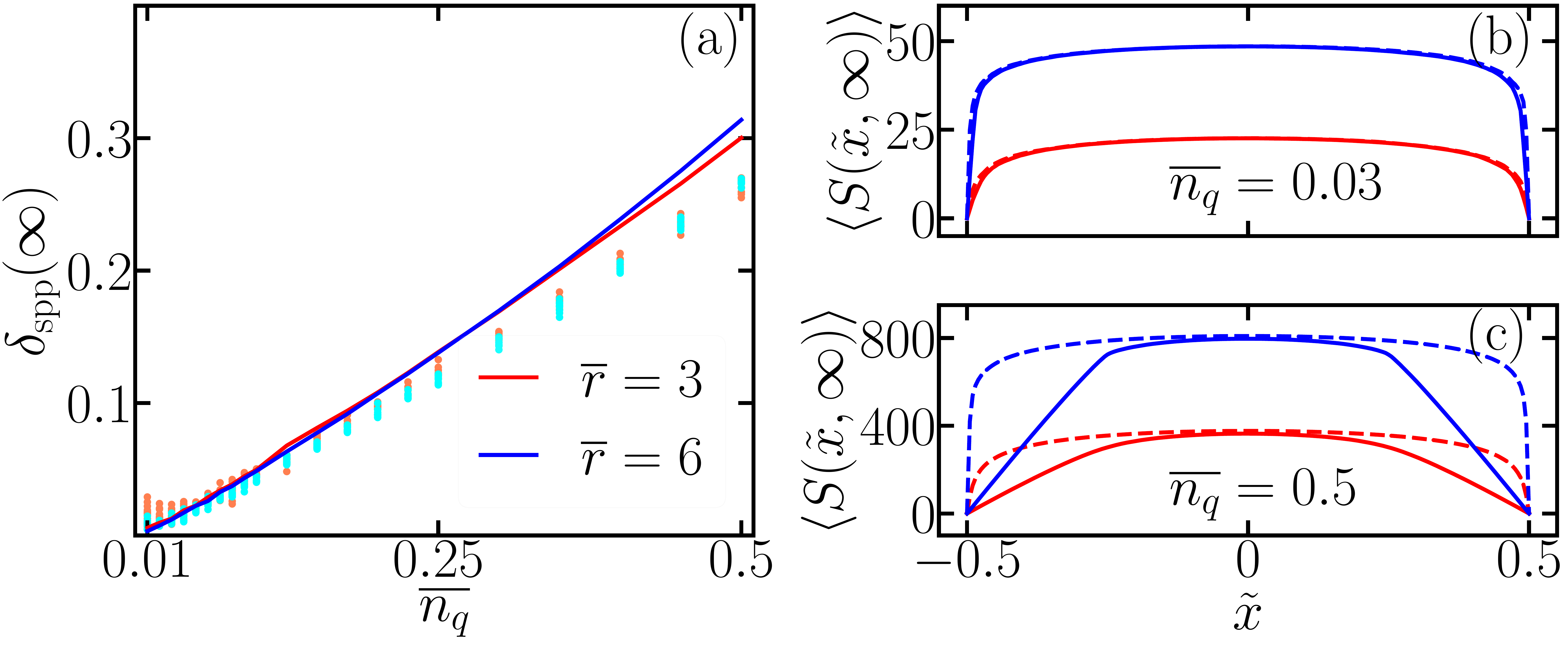}
    \caption{(a) Relative 1-norm of the difference between real and superposition Page curves in a system of $M = 200$ modes. Solid lines represent the results of circuits with identical $r_i = \overline{r}$ squeezers, while scatter dots show results of circuit where random squeezers $r_i$ uniform in a range $[0.5\overline{r}, 1.5\overline{r}]$ with fixed total squeezing strength $N_q \overline{r}$. Page curves with $\overline{n_q} = 0.03$ and $\overline{n_q} = 0.5$ are shown in (b) and (c). Red and blue lines correspond to $\overline{r} = 3$ and $\overline{r} = 6$. Solid and dashed lines represent real and superposition results.
    \label{fig:static_spp}
    }
\end{figure}

In terms of Page curves, as we explained in Sec.~\ref{sec:equi_solu_general}, the graph topology is irrelevant as the entire dynamics is equivalent to a passive, Haar-random, global unitary; therefore, we can simply consider the 1-D case. Furthermore, a Gaussian unitary can be decomposed into a layer of squeezers concatenated by passive unitaries (Euler decomposition, see Appendix~\ref{app:random_mat}), we can effectively push all squeezers to the first step and then apply a global Haar random passive linear transform (see Fig.~\ref{fig:schematic_Page}). Therefore, each Page curve is characterized by a list of squeezing strength $\{r_k\}_{k=1}^{N_q}$.  In this case, as long as $N_q\ll M$, we can regard the squeezers as sparse. We denote the average squeezing strength as $\overline{r}=\sum_{\ell=1}^{N_q} r_\ell/N_q$ and the density of squeezers $\overline{n_q}=N_q/M$. When $N_q\ll |\calL|$, superposition $\braket{S(\calL,\infty)}\simeq  \braket{S_{\rm spp}(\calL,\infty)}$ holds, with each single-squeezer result given in Eq.~(\ref{SL_page}). In the large squeezing limit, we can further utilize Eq.~(\ref{SL_page_sim}) to obtain
\begin{align}
&S\left(\calL,\infty\right)
\nonumber
\\
&\simeq M \overline{n_q}\left\{\frac{1}{2}\log_2\left[\frac{|\calL|}{|\calG|}\left(1-\frac{|\calL|}{|\calG|}\right)\right]+\frac{1}{\ln2}\left(\overline{r}+1\right)-1\right\},
\label{SL_sim_page}
\end{align}
We see a dependence on statistical quantities $\overline{r}, \overline{n_q}$, while the shape of the curve is invariant in the bulk at large squeezing limit, as can be seen in Fig.~\ref{fig:static_spp} (b) and (c).

We numerically examine the validity of the linear superposition in Page curves via the relative deviations $\delta_{\rm spp}(t)$ in Fig.~\ref{fig:static_spp}. Indeed when the squeezer density is low, we see a good agreement, as shown in Fig.~\ref{fig:static_spp} (b); while when the squeezers are dense, substantial deviation can be found, as shown in Fig.~\ref{fig:static_spp} (c). The transition is captured by the relative deviation $\delta_{\rm spp}(\infty)$ in Fig.~\ref{fig:static_spp} (a), where $\delta_{\rm spp}(\infty)$ increases linearly with the squeezer density $\overline{n_q}$.

Although when $\overline{n_q}$ is not small, superposition does not hold, we can further numerically explore the Page curves' dependence on the parameters. To consider the typical case, we randomly generate $\{r_k\}_{k=1}^{N_q}$ and plot the normalized Page curves in Fig.~\ref{fig:pages}(b) for different system sizes. We find that all Page curves coincide as long as $\overline{r}, \overline{n_q}$ are the same; and independent of $M$ when $M$ is large Similar to the single-squeezer case. Therefore, we can plot the case with identical squeezer $\overline{r}$ as a benchmark. To demonstrate the coincidence, we consider random cases with $r_i \in [0.8\overline{r}, 1.2\overline{r}], 1\le i \le N_q$, while guaranteeing the average equaling $\overline{r}$~\cite{Note2}. Indeed, all the random points lie right on top of the benchmark of uniform squeezing with the same $\overline{r}, \overline{n_q}$. Since $\overline{r}$ and $\overline{n_q}$ are both scale-free statistical quantities, this indicates a well-defined continuum limit of CV Page curves.



\subsection{Dynamics with linear superposition}
\label{sec:superposition}

\begin{figure}
    \centering
    \includegraphics[width = 0.48\textwidth]{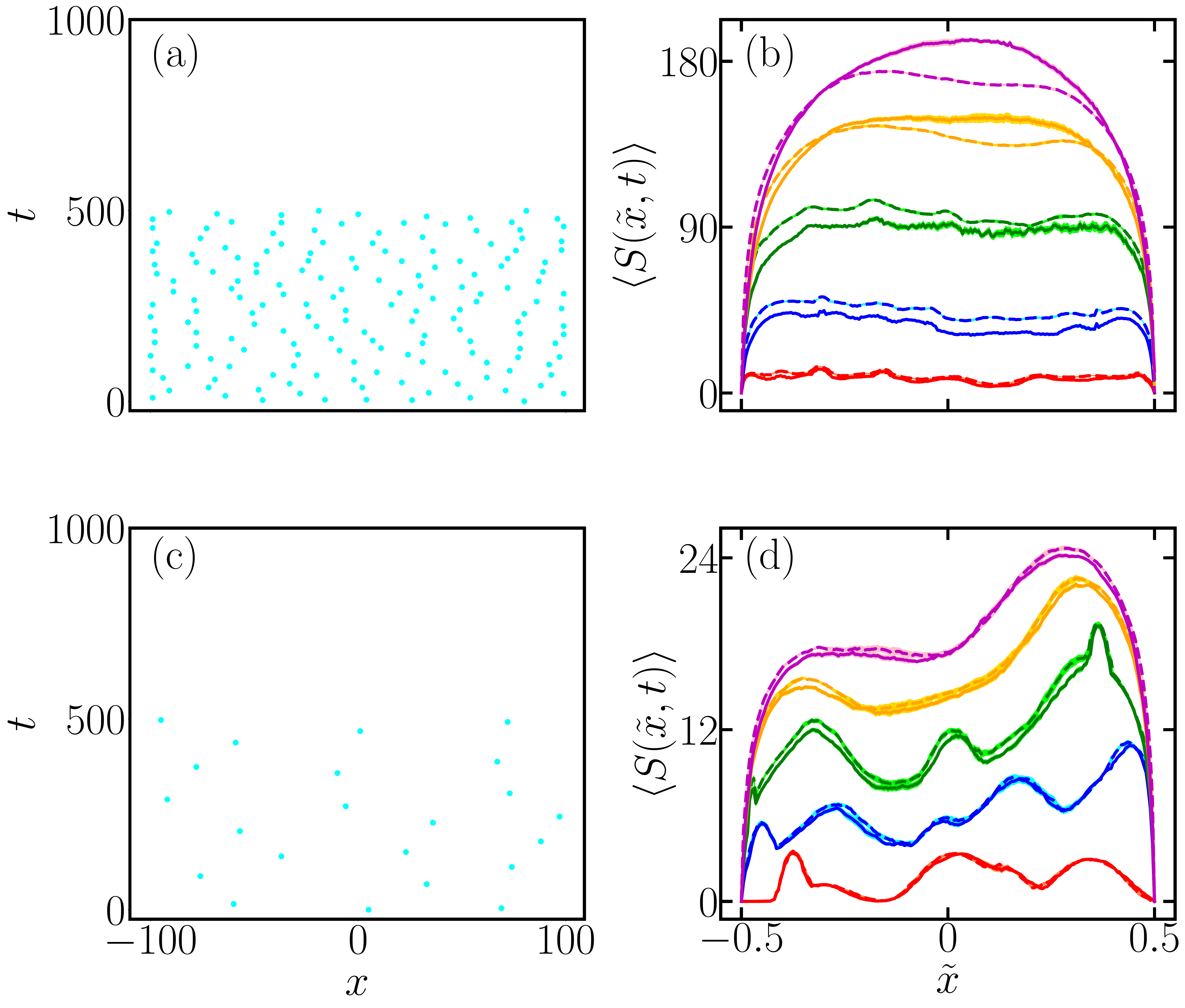}
    \caption{Examples for linear superposition of random squeezers. Squeezing strength of each random squeezers $r_i$ is uniform in the range $[1,3]$. In the left panel, Cyan dots represent the places of squeezers in space-time, when the minimum distance $d=20$ (a) and $d=60$ (c). In the right panel, (b) and (d) are Page curves for the setup with $d=20$ and $d=60$ separately. In each figure, lines from bottom to top show the Page curve at time $t=100, 300, 500, 700, 900$. Solid lines represent average original Page curves and dashed lines are average superposition ones. Shaded area with light colors shows the precision. 
    \label{fig:sparse_dense}
    }
\end{figure}

Fig.~\ref{fig:dynamic_spacetime_spp_N3} already confirms superposition through the entire evolution for a simple case, to verify it in a more general setting, we consider circuits with squeezers randomly distributed in space-time. Guided by the theory in Sec.~\ref{sec:spp_theory}, we expect the minimum distance between the squeezers to be the dominating factor of the deviation $\Delta S_{\rm spp}(t)$; However, simple random sampling methods inevitably lead to the appearance of clusters, where squeezers can be close to each other. In order to tune the minimum distance $d$, while preserving the random nature of the circuit set-up so that the results apply in general, we adopt the random Poisson sampling method~\cite{bridson2007} to control the space-time distances between the squeezers. 


With the squeezers randomly chosen, we evaluate the entanglement dynamics for circuits with random squeezers of different minimum distances. Two examples are given in Fig.~\ref{fig:sparse_dense}: When the squeezers are sparse ($d=60$ is large) as indicated by the cyan dots in Fig.~\ref{fig:sparse_dense}(c), the superposition results $\braket{S_{\rm spp}(\tilde{x},t)}$ (dashed lines) agree well with the true values $\braket{S(\tilde{x},t)}$ (solid lines) at various time steps in Fig.~\ref{fig:sparse_dense}(d); when the squeezers are dense ($d$ is small), substantial deviations from the linear superposition can be observed, as demonstrated in Fig.~\ref{fig:sparse_dense}(a)(b) for the case of $d=20$. 

\begin{figure}
    \centering
    \includegraphics[width = 0.3\textwidth]{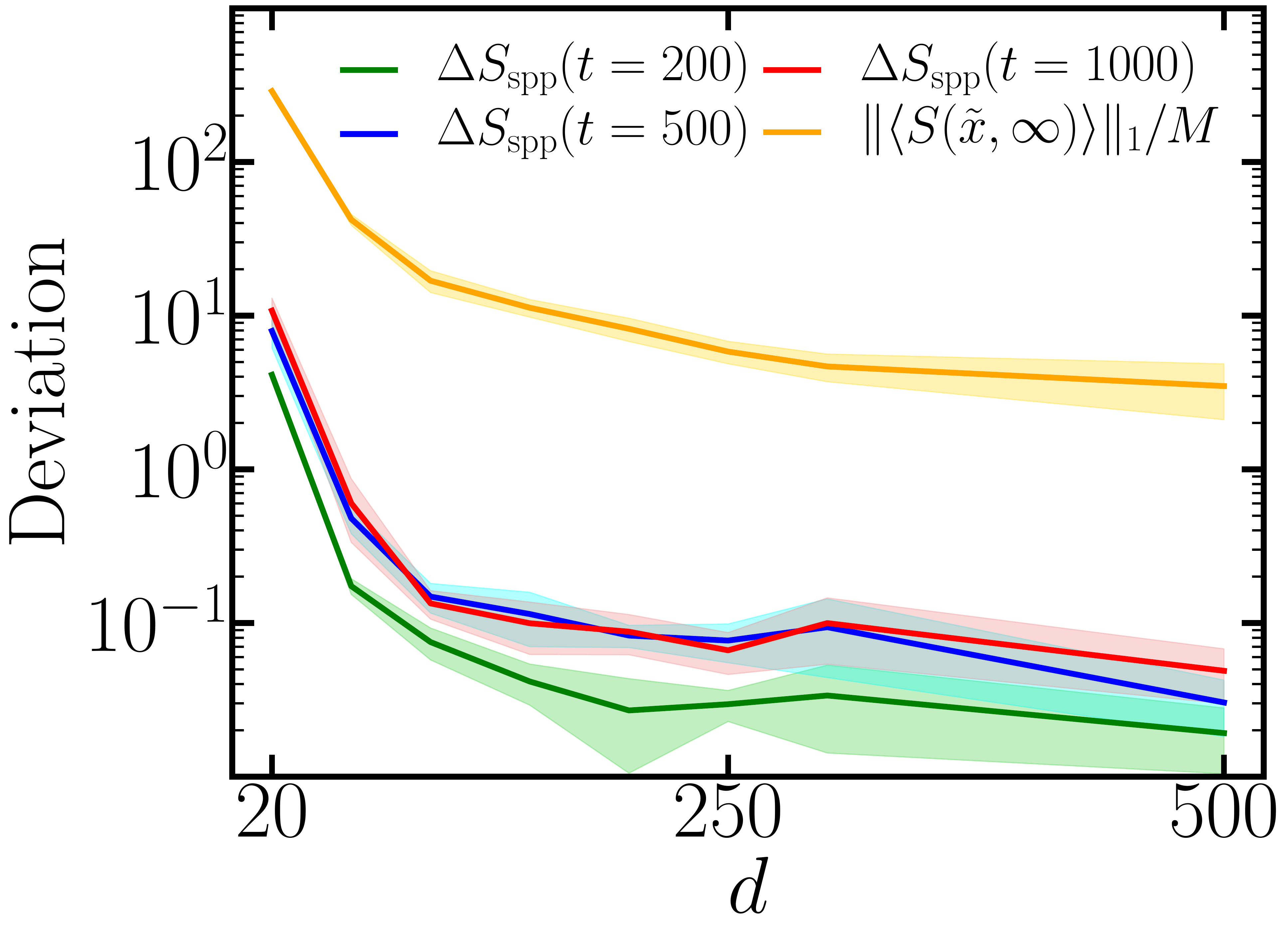} 
    \caption{Absolute deviation (per mode) of superposition measured by $\Delta S_{\rm spp}(t)$, with different minimum distances between random squeezers $d$. Time $t$ is fixed at early ($t=200$, green), medium ($t=500$, blue) and late time ($t=1000$, red). The shaded areas show the fluctuations in different squeezer space-time configurations. In comparison, the entanglement per mode $\|\braket{S(\tilde{x},\infty)}\|_1/M$, averaged over configurations with fixed $d$ are plotted in orange.
    System size $M=200$. Squeezers with random strength $r_i\in[1,3]$ are randomly distributed before $t=500$ following Poisson disk sampling, similar to Fig.~\ref{fig:sparse_dense}. 
    \label{fig:sqs_spp_rn}
    }
\end{figure}

To systematically examine the transition from `dense' to `sparse' squeezers, we calculate the deviation $\Delta_{\rm spp}(t)$ for distances $0\le d\le 500$ at various time steps in Fig.~\ref{fig:sqs_spp_rn}, where the deviation decreases monotonically with $d$ up to numerical precision. Although for a smaller $d$, the deviation is larger, the relative deviation is still below $10\%$ of the static value $\|\braket{S(\tilde{x},\infty)}\|_1/M$ (orange). Therefore, we conclude that in generic CV quantum networks with sparse squeezers, linear superposition of entanglement growth holds.

\section{Discussion} 
In this paper, we reveal a mapping between entanglement formation dynamics in random CV networks to random walk on general graphs. This mapping allows analytical solutions of the entanglement entropy dynamics, Page curves and scrambling time for an arbitrary network topology. On networks respecting locality, the solution enables the understanding of three unique features of entanglement formation dynamics---an entanglement light cone, an entanglement sudden-growth period and parameter-dependent Page curves. Our results have implications in quantum network protocol design, e.g., the entanglement light cone will place bounds on the latency in entanglement distribution, and also on the fundamental understanding of many-body systems. Lastly, let us point out some future directions: it will be interesting to extend the entanglement light cone to long-range interacting systems; extending the multipartite entanglement witness to more general input states will bring further insights of quantum entanglement; exploration of the connection between entanglement dynamics with statistical properties of random networks such as connectivity distribution will lead to a full statistical theory of CV quantum networks.

\begin{acknowledgements}
This work is supported by Army Research Office under Grant No. W911NF-19-1-0418 and the University of Arizona. Q.Z. thanks Norman Yao and Beni Yoshida, Saikat Guha, Bihui Zhu for discussions.
\end{acknowledgements}

\begin{appendix}

\section{Details of Gaussian unitaries and random matrices}
\label{app:random_mat}

A zero-mean general Gaussian unitary $U_{\bm S}$ is specified by a symplectic matrix $\bm S$, which can be decomposed into a product of `passive linear optics' operations and `squeezing' operations (Euler decomposition),
\be
U_{\bm S}=U_{\bm K} U_{\bm S(\left\{r_k\right\})} U_{\bm L},
\label{Euler_decomp}
\ee 
where $\bm K, \bm L$ are symplectic orthogonal, and correspond to beamsplitters and phase-shifters. These two components are often named `passive linear optics' since they preserve the mean photon number. Single-mode squeezing operations, which changes photon number, are characterized by their strengths $r_k$ and represented by the diagonal symplectic matrix $\bm S(\{r_k\}) = \bigoplus_k \text{Diag}\left( e^{-r_k}, e^{r_k} \right)$, where `$\bigoplus$' denotes the direct sum. 
An additional property of zero-mean Gaussian unitary is that concatenating multiple Gaussian gates still produces a Gaussian gate, i.e., $U_{\bm S_1} U_{\bm S_2}=U_{\bm S_1 \bm S_2}$.

First, we specify a Haar random two-mode passive linear transform parameterized by angles $\theta$, $\phi_1, \phi_2$ and $\phi_3,\phi_4$, as shown below.
\begin{figure}[H]
    \centering
    \includegraphics[width = 0.15\textwidth]{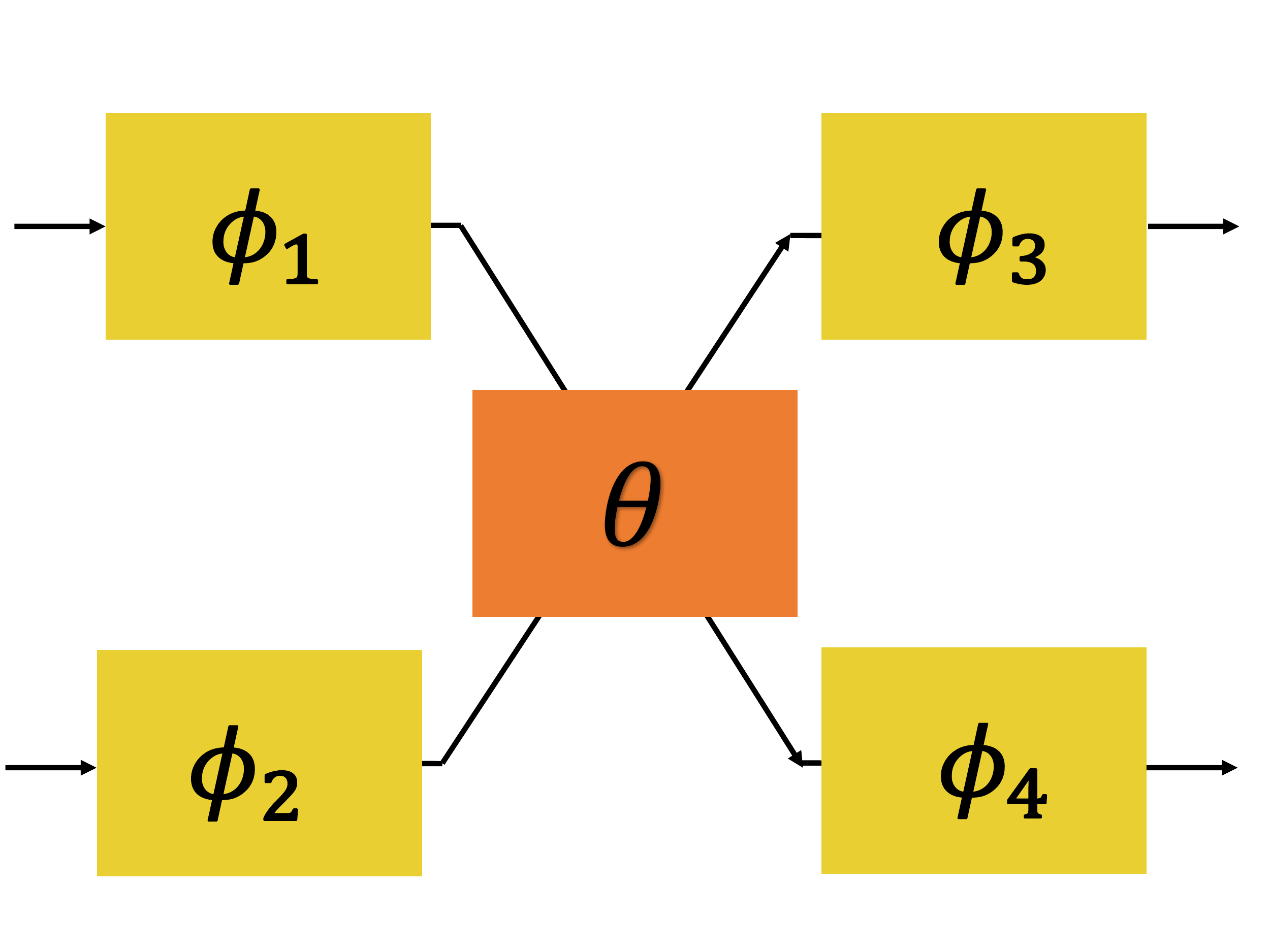} 
\end{figure}
\noindent It consists of a beamsplitter described by the unitary $\exp\left[\theta \left(\hat{a}\hat{b}^{\dagger} - \hat{a}^{\dagger}\hat{b}\right)\right]$ on input modes $a,b$, where $\cos^2(\theta)$ is the transmissivity of the beamsplitter and four additional phase shifters, $R(\phi_1)$, $R(\phi_2)$, $R(\phi_3)$ and $R(\phi_4)$, two of which are ahead of beamsplitters and the others after it. The unitary of a phase shifter with amount $\phi$ on mode $\hat{a}$ is described as $R(\phi) = \exp\left(i\phi\hat{a}\hat{a}^{\dagger}\right)$. The corresponding unitary matrix and symplectic matrix are respectively
\begin{align}
    &\bm U_{B} = \begin{pmatrix}
    e^{i(\phi_1 + \phi_3)}\cos\theta & -e^{i(\phi_2 + \phi_3)}\sin\theta \\
    e^{i(\phi_1+\phi_4)}\sin\theta & e^{i(\phi_2+\phi_4)}\cos\theta
    \end{pmatrix}, 
    \label{eq:bs_unitary}
    \\
    &\bm S_{B} = \bm P^{T} \begin{pmatrix}
    \operatorname{Re}(\bm U_{B}) & -\operatorname{Im}(\bm U_B)\\
    \operatorname{Im}(\bm U_B) & \operatorname{Re}(\bm U_{B})
    \end{pmatrix} \bm P.
    \label{eq:bs_symplectic}
\end{align}
Here the $4\times 4$ permutation matrix $\bm P$ exchanges the second and third item, i.e., $\bm P=\bm I_{14}\bigoplus \bm X_{23}$, with Pauli matrices $\bm I, \bm X$.
The notations $\operatorname{Re}(.), \operatorname{Im}(.)$ denote the real and imaginary part of a complex matrix. To ensure Haar random of Eq.~(\ref{eq:bs_unitary}), all phase angles $\phi_1,\phi_2,\phi_3, \phi_4$ are chosen uniform in $[0,2\pi)$ and $\cos^2\theta$ uniform in $[0, 1]$. Here we also reveal a one-to-one correspondence between orthogonal symplectic matrix and unitary matrix through Eq.~(\ref{eq:bs_symplectic}) (and its generalization to more modes), which allows us to sample multi-mode Haar random passive linear transform from standard algorithms in Ref.~\cite{mezzadri2006} directly.

\section{Entanglement entropy}
\label{app:entropy}

The entanglement entropy $S(\calL,t)$ can be efficiently calculated through keeping track of the covariance matrix of $\calL$ (Equivalently, one can consider the covariance matrix of subsystem ${\protect \calR}$.)
\be 
\bm V_{ij}=\braket{\{\bm X_i,\bm X_j\}}, i,j \in \calL,
\ee 
where $\{,\}$ is the anti-commutator and $\bm X_i,\bm X_j$ are components of quadrature operators $\bm X$ that corresponds to subsystem $\calL$. Under the dynamics of each local unitary $U_{\bm S}$, the covariance matrix evolves as $\bm V\to \bm S \bm V \bm S^T$. From the symplectic eigenvalues~\cite{Weedbrook_2012} $\{\nu_i, 1\le i \le |\calL|\}$ of $\bm V$, we can obtain the von Neumann entropy of $\calL$ (which equals that of $\calR$ due to purity of the global system)
\be 
S\left(\calL,t\right)=S(\calR,t)=\sum_{i=1}^{|\calL|} g(\left(\nu_i-1\right)/2),
\ee
where each term
$
g(x)  = (x+1)\log_{2}(x+1) - x\log_{2}\left(x\right)
$
is the entropy of a thermal state with mean photon number $x$. 
Alternatively, we can also choose Renyi entropy of the order $\alpha$, which can be calculated as~\cite{Camilo2019}
\be 
S_{\alpha}(\calL,t)  = 
\frac{1}{\alpha-1} \sum_{i=1}^{|\calL|} \log_2{g_{\alpha}((\nu_i-1)/2)},
\label{eq:Renyi entropy}
\ee
where the $g_{\alpha}\left(\cdot\right)$ is defined as $g_{\alpha}(x) = \left(x+1\right)^{\alpha} - \left(x\right)^\alpha$

Combining the above, when there is a single non-zero symplectic eigenvalue, we can write a single function $g(x)$, as we have done in Eq.~(\ref{SL_preliminary}): for von Neumann entropy $g(x)= (x+1)\log_{2}(x+1) - x\log_{2}\left(x\right)$ and for Renyi entropy $g(x)=\log_2\left[(x+1)^\alpha-x^\alpha\right]/(\alpha-1)$.

In the large squeezing limit, to the first order, we have $S\left(\calL,t\right)= S(\calR,t)\simeq  \log_2(\prod_{i=1}^{|\calL|}\nu_i)=\log_2 (\sqrt{\det \bm V})$, which equals a projected phase space volume similar to the one identified through average OTOCs in Ref.~\cite{zhuang2019scrambling}. The purity of the entire system also guarantees a conservation of the entire phase space volume, analog to the ``Quantum Liouville’s theorem'' in Ref.~\cite{zhuang2019scrambling}. This interpretation also holds for Renyi entropy: e.g. for the Renyi-2 entropy $S_{2}(\calL)=S_{2}(\calR)=\log_2(\sqrt{\det \bm V})$, which can be interpreted as the logarithmic volume; For other choices of $\alpha\neq 1$, when $\nu_i$'s are large, we still have $S_\alpha(\rho)\simeq \log_2(\sqrt{\det \bm V})$. In this sense, the entanglement growth dynamics can be connected to projected phase-space volume growth (also see Ref.~\cite{lerose2020}), and it is clear that any choices of entanglement entropy will have similar dynamics.

Note that here Renyi entropy is not directly connected with t-designs~\cite{liu2018generalized}, since the symplectic eigenvalues are related to eigenvalues of $i\bm\Omega\bm V$, which are not simple polynomials in general. Here $\bm \Omega=\bigoplus_{k=1}^N -i\bm Y$ is the symplectic metric.

It is worthy to comment that the diffusive growth of entanglement identified in Sec.~\ref{sec:light_cone} does not contradict the known result that good approximations ($t$-design) of the global Haar random unitary only requires a random circuit with the number of layers linear in system size~\cite{harrow2018approximate}. Mathematically, in a CV system it is the covariance matrix $\bm V$ that is transformed as $\bm V\to \bm S \bm V \bm S^T $ under the Haar random matrix $\bm S$. Both the von Neumann and Renyi entanglement entropies are given by the symplectic eigenvalues of $\bm V$, which is not a simple polynomial of $\bm S$. This contrasts with the case in DV systems, where the random matrices act on the density matrix directly and Renyi entropies are simple polynomials of them.

\section{Details of the random-walk mapping} 
\label{app:random_derivation}
In terms of the weights, consider two modes $\bm x, \bm x^\prime$ on a connected edge
\begin{subequations}
\label{bs1}
\begin{align}
a_{\bm x,t}&=e^{i\theta_{\bm x,t}}\sqrt{w_{\bm x,t}} a_{\rm SV}+{\rm vac},
\\
a_{\bm x^\prime,t}&=e^{i\theta_{\bm x^\prime,t}}\sqrt{w_{\bm x^\prime,t}} a_{\rm SV}+{\rm vac}.
\end{align}
\end{subequations}
A general passive gate $U(t,\bm x,\bm x^\prime)$, with transmissivity $\tau$, on modes $\bm x,\bm x^\prime$ would lead to the evolution
\begin{subequations}
\label{bs2}
\begin{align}
&a_{\bm x,t+1}=
e^{i\theta_{\bm x,t+1}^\prime}\left(\sqrt{\tau}\sqrt{w_{\bm x,t}}+\sqrt{1-\tau} e^{i\theta}\sqrt{w_{\bm x^\prime,t}}\right) a_{\rm SV},
\\
&a_{\bm x^\prime,t+1}=
e^{i\theta_{\bm x^\prime,t+1}^\prime}\left(\sqrt{1-\tau} \sqrt{w_{\bm x,t}}-\sqrt{\tau} e^{i\theta}\sqrt{w_{\bm x^\prime,t}}\right) a_{\rm SV},
\end{align}
\end{subequations}
where the angles $\theta=(\theta_{\bm x^\prime,t}-\theta_{\bm x,t}), \theta_{\bm x,t+1}^\prime,\theta_{\bm x^\prime,t+1}^\prime$ are uniform in $[0,2\pi)$ and we left out the vacuum terms. Therefore we can write the new modes in the same form as Eq.~(\ref{axt}) with
\begin{subequations}
\label{bs3}
\ba 
w_{\bm x,t+1}&=&\tau w_{\bm x,t}+(1-\tau)w_{\bm x^\prime,t}
\nonumber
\\
&&
+2\sqrt{\tau(1-\tau)}\sqrt{w_{\bm x,t}w_{\bm x^\prime,t}}\cos\theta,
\\
w_{\bm x^\prime,t+1}&=&(1-\tau) w_{\bm x,t}+\tau w_{\bm x^\prime,t}
\nonumber
\\
&&
-2\sqrt{\tau(1-\tau)}\sqrt{w_{\bm x,t}w_{\bm x^\prime,t}}\cos\theta,
\ea 
\end{subequations}
and the angles $\theta_{\bm x,t+1},\theta_{\bm x^\prime,t+1}$ uniformly random in $[0,2\pi)$.
We can obtain the ensemble averaged (over the gates $U(t,\bm x,\bm x^\prime)$, with transmissivities $\tau$ uniform in $[0,1)$) evolution as
\be 
\braket{w_{\bm x,t+1}}=\braket{w_{\bm x^\prime,t+1}}=\frac{1}{2}\left(\braket{w_{\bm x,t}}+\braket{w_{\bm x^\prime,t}}\right).
\ee 
This describes a symmetric random walk along the edge $\overline{\bm x\bm x^\prime}$.

\section{Calculation of the variance of $\eta_{\calL,\infty}$ at equilibrium} 
\label{app:eta_var}

The weights $\{w_{\bm x,t}\}$ determine the entanglement, essentially they can be considered as the amplitude squared of a random complex number $\alpha_{\bm x,t}$, i.e. $w_{\bm x,t}=|\alpha_{\bm x,t}|^2$. Normalization requires $\sum_{\bm x\in \calG} |\alpha_{\bm x,t}|^2=1$. Due to the Haar randomness at $t=\infty$, we assume that this is the only constraint, therefore $\alpha_{\bm x,t}$'s are random complex numbers on a high dimension sphere. We can calculate the distribution of weights $\eta_{\calL,\infty}$ through
\begin{align}
&P(\eta_{\calL,\infty}=\eta)=
\nonumber
\\
&\frac{\int \prod_{\bm x\in \calG}d^2\alpha_{\bm x,t}\  \delta\left(\eta-\sum_{\bm x\in \calL} |\alpha_{\bm x,t}|^2\right) \delta\left(1-\sum_{\bm x\in \calG} |\alpha_{\bm x,t}|^2\right)}{\int \prod_{\bm x\in \calG}d^2\alpha_{\bm x,t} \ \delta\left(1-\sum_{\bm x\in \calG} |\alpha_{\bm x,t}|^2\right)}
\\
&=\frac{\int \prod_{\bm x\in \calR}d^2\alpha_{\bm x,t}\  S_{2|\calL|-1}(\sqrt{\eta}) \delta\left(1-\eta-\sum_{\bm x\in \calR} |\alpha_{\bm x,t}|^2\right)}{\sqrt{\eta} S_{2|\calG|-1}(1)}
\\
&=\frac{ S_{2|\calL|-1}(\sqrt{\eta}) S_{2|\calR|-1}(\sqrt{1-\eta})}{2\sqrt{\eta}\sqrt{1-\eta}S_{2|\calG|-1}(1)}
\\
&=\frac{\Gamma(|\calG|)}{\Gamma(|\calL|)\Gamma(|\calR|)}\eta^{|\calL|-1}(1-\eta)^{|\calR|-1}.
\label{P_eta}
\end{align}
Here we utilized the $N$-dimensional sphere area formula
\begin{align}
&S_N(R)\equiv \int \prod_{\ell=1}^N dx_\ell \ \delta\left(R-\sqrt{\sum_{\ell=1}^{N+1} x_\ell^2}\right)
\\
&=2R\int \prod_{\ell=1}^N dx_\ell \ \delta\left(R^2-\sum_{\ell=1}^{N+1} x_\ell^2\right)=\frac{2\pi^{\frac{N+1}{2}}}{\Gamma(\frac{N+1}{2})}R^N,
\end{align}
where we have used $\delta(x^2-a^2)=(\delta(x-a)+\delta(x+a))/(2a)$. One can easily verify that Eq.~(\ref{P_eta}) is normalized and $\braket{\eta_{\calL,\infty}}=|\calL|/|\calG|$. Moreover, the fluctuation can be obtained as
\be 
{\rm var}(\eta_{\calL,\infty}) = \frac{|\calL||\calR|}{|\calG|^2(|\calG|+1)}.
\ee

\section{Continuum limit of the multiple-squeezer case}
\label{app:conti_multi}
\begin{figure}
    \centering
    \includegraphics[width = 0.45\textwidth]{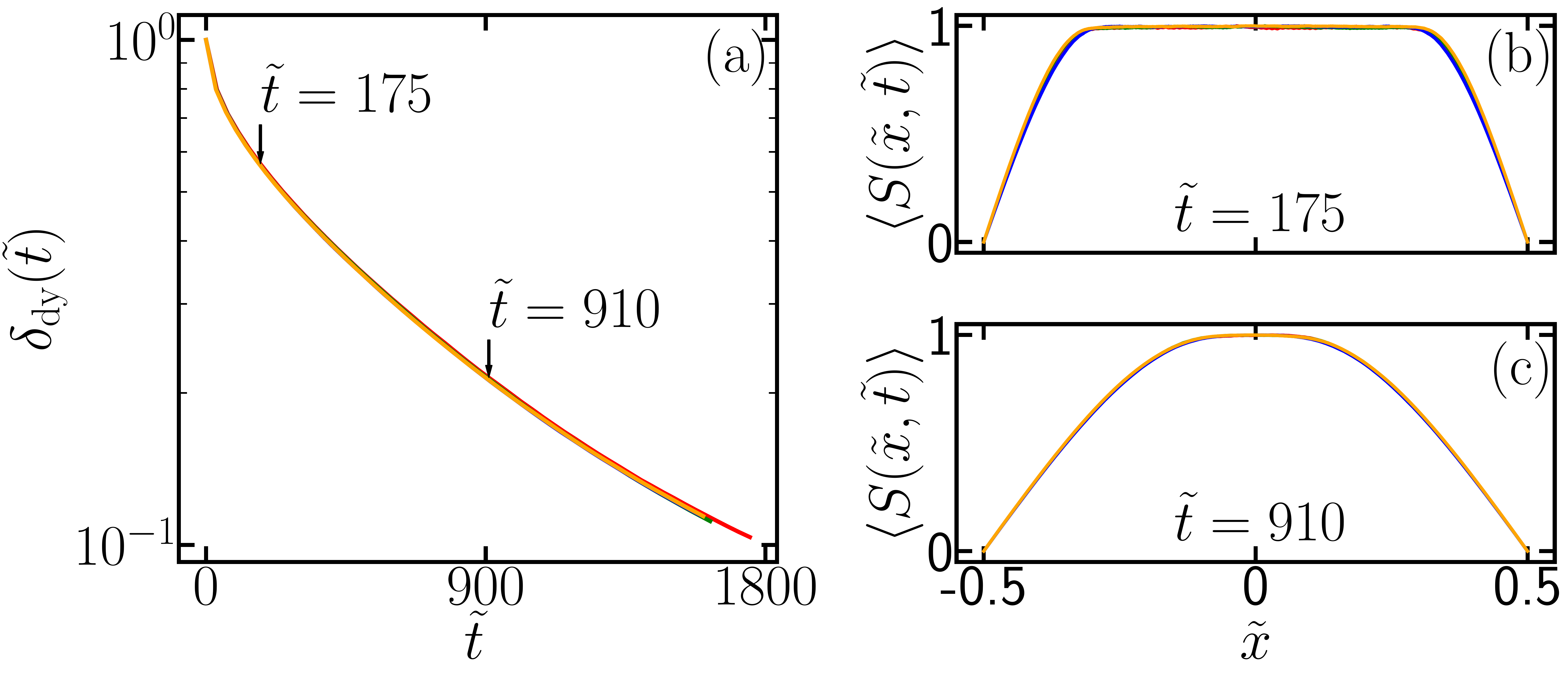}
    \caption{Relative 1-norm deviation $\delta_{\rm dy}(\tilde{t})$ for dynamic entropy curves in 1-D Cartesian graph. One layer of $N_q=M$ squeezers is applied at $t=0$ with an identical strength $r=5$. All curves are re-scaled and shown in the same way as in Fig.~\ref{fig:dynamic_scale_invariance}.
    \label{fig:sqs_log(delta)2}
    }
\end{figure}

In this section, we show that the same continuum limit, identified in Sec.~\ref{sec:continuum} (see Fig.~\ref{fig:dynamic_scale_invariance}) for a single-squeezer, still holds when there are multiple squeezers. In Fig.~\ref{fig:sqs_log(delta)2}, one layer of $N_q=M$ squeezers are applied in the first time step, and we still see a good agreement between the relative 1-norm (Fig.~\ref{fig:sqs_log(delta)2}(a)) at different times. Snapshots of entanglement entropy curve at various times also overlap entirely (Fig.~\ref{fig:sqs_log(delta)2}(b), (c)), after a re-scaling of space-time. 
Note that in this case, even with a dense squeezer distribution where the superposition principle does not hold anymore, this continuum limit still holds.

\section{Phenomenological model: epidemiology with diffusion}
\label{app:epidemiology}

Although the dynamics of $\braket{S(x,t)}$ can be captured by the diffusion of weights and Eq.~(\ref{SL}) that connects weights to entanglement entropy. Alternative models are possible to capture major phenomena in Sec.~\ref{sec:light_cone} for the 1-D Cartesian graph. We notice that in our CV circuits, squeezing behaves as the source of entanglement generation~\cite{tserkis2020}; and while it diffusively spreads out, its strength at each mode also decays due to the effective loss during interactions with other modes; therefore we introduce a field $G(x,t)$ to model this diffusive source that triggers entanglement growth and device the following set of coupled diffusion-growth equations to give a theory prediction $S_T\left(x,t\right)$ for the ensemble averaged entanglement entropy $\braket{S\left(x,t\right)}$ 
\begin{subequations}
\label{couple_KPZ}
\begin{align}
&\partial_{t}S_T(x,t) = AG(x,t)\left[c - S_T\left(x,t\right)\right],
\label{infection}
\\
&\partial_{t}G(x,t) = D\partial^2_{x} G(x,t) + fG(x,t)^{2},
\label{KPZ}
\end{align}
\end{subequations}
\begin{figure}
    \centering
    \includegraphics[width = 0.45\textwidth]{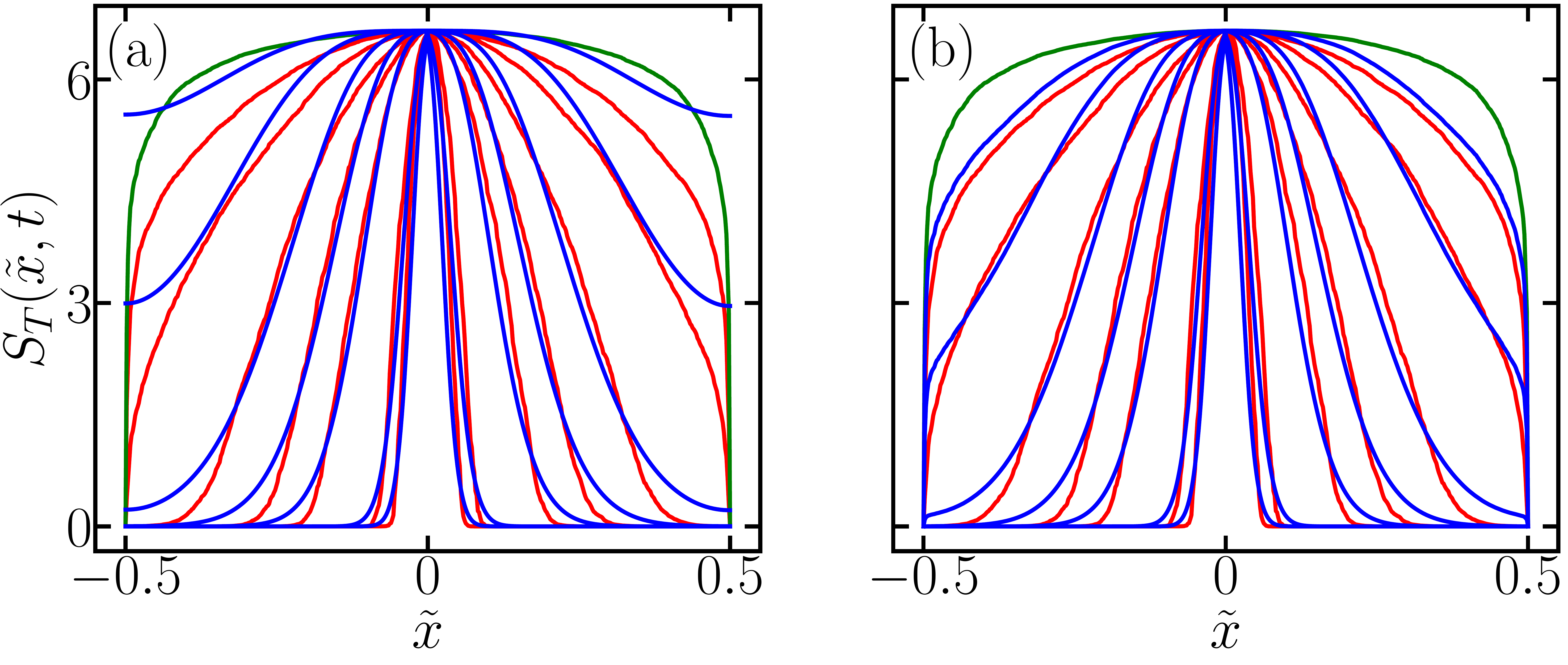}
    \caption{Coupled diffusion-growth model for the single squeezer circuit. The squeezer with $r=5$ is placed at the center of a system consisting of $M=501$ modes. Parameters for the model is chosen to be $D = 2, f = -0.7, A = 2.3$. Dynamic curves from inside to outside correspond to entanglement entropy at time $t = 50, 100, 200, 500, 1000, 2000, 5000, 10000$. Red and blue lines represent real and the coupled diffusion-growth model dynamic entropy evolution. The Page curve $\braket{S(x,\infty)}$ shown by the green line is given as a reference. (a) $c=S_0(r)$. (b) $c=\braket{S\left(x, \infty\right)}$. 
    \label{fig:model}
    }
\end{figure}
where $A, c, D, f$ are four constants. Eq.~(\ref{KPZ}) describes a nonlinear diffusion process of the `source', where the nonlinear term $fG{\left(x,t\right)}^2$ corrects the early stage dynamics. Eq.~(\ref{infection}) describes an `infectious' saturation process with a growth rate proportional to `uninfected population', $c - S_T\left(x,t\right)$, triggered by the source $G(x,t)$. We note that both nonlinear diffusion equations~\cite{nahum2017quantum} and epidemiology models~\cite{qi2018quantum} have been separately used in modeling quantum information scrambling. This model shows an interesting combination of both to describe a unique CV entanglement growth process.

The initial condition for the source field $G\left(x,t\right)$ is a delta-function at the squeezer's position, and uniform zero for the entanglement $S_T\left(x,t\right)$. We choose $G\left(x,t\right)$ obeying the von Neumann boundary condition (reflection boundary conditions) such that $\partial_{x} G(\pm N,t) = 0$. It is straightforward to check that $S_T\left(x,t\right)=c$ is a steady state solution of Eqs.~(\ref{couple_KPZ}). Considering the steady state of the real entanglement entropy in Fig.~\ref{fig:pages_1D}, we expect Eqs.~(\ref{couple_KPZ}) to only describe the evolution of entanglement in the bulk. Indeed, when we numerically solve Eqs.~(\ref{couple_KPZ}) in a system with $M=501$ modes and fix the steady state $c$ to be the maximal height $S_0(r)$, we find good agreement before boundary effects come in at late time (see Fig.~\ref{fig:model}(a)). In particular, this model does not guarantee zero entanglement entropy on the boundary.

Therefore, Eqs.~(\ref{couple_KPZ}) will be able to capture the entanglement growth of the bulk, before boundary effects become important. More precise models can be constructed by further fine-tuning Eqs.~(\ref{couple_KPZ}) to different boundaries. As an example, we can implement a position-dependent $c=\braket{S\left(x,\infty\right)}$, such that steady state is consistent with the Page curve. Then, we numerically solve Eqs.~(\ref{couple_KPZ}) and find the theory prediction in a nice agreement with the real data from random circuit simulations in Fig.~\ref{fig:model}(b). 


\end{appendix}



%

\end{document}